\documentclass[preprint,12pt]{elsarticle}

\usepackage{amssymb}
\usepackage{amsthm}
\usepackage{setspace}
\usepackage{lineno}
\usepackage{listings}
\usepackage{color}
\usepackage[dvipsnames,svgnames,x11names]{xcolor}
\usepackage[hang,flushmargin]{footmisc}
\usepackage{orcidlink}
\usepackage{xspace} 
\usepackage{tikz}
\usetikzlibrary{shapes,arrows}
\usepackage{nicematrix}
\usepackage{comment}
\usepackage{cancel}
\usepackage{soul}
\usepackage{nicematrix}
\usepackage{subcaption}
\usepackage{hyperref}
\newcommand{\highlight}[2][yellow]{\mathchoice%
  {\colorbox{#1}{$\displaystyle#2$}}%
  {\colorbox{#1}{$\textstyle#2$}}%
  {\colorbox{#1}{$\scriptstyle#2$}}%
  {\colorbox{#1}{$\scriptscriptstyle#2$}}}%

\lstdefinestyle{customcpp}{
    language=C++,
    basicstyle=\ttfamily\footnotesize,  
    keywordstyle=\bfseries,             
    commentstyle=\itshape\color{gray},  
    stringstyle=\color{black},            
    numbers=left,                        
    numberstyle=\tiny\color{gray},      
    stepnumber=1,                        
    frame=single,                        
    captionpos=b,                        
    breaklines=true,                     
    tabsize=4,                           
    morekeywords={label, forAll, const, autoPtr} 
}

\def\figureautorefname{Fig.}
\def\equationautorefname{Eqn.}

\newcounter{bla}

\journal{}

\begin{document}

\begin{frontmatter}



\title{An advanced fully-implicit solver for heterogeneous porous media based on foam-extend}




\author[a]{Roberto Lange}
\author[a]{Gabriel M. Magalhães}
\author[a]{Franciane F. Rocha\corref{author}}
\author[a]{Hélio Ribeiro Neto}

\cortext[author] {Corresponding author.\\\textit{E-mail address:} franciane.rocha@wikki.com.br}
\address[a]{WIKKI Brasil LTDA, Rua Aloísio Teixeira, 278, Prédio 3, sala 301, Ilha da Cidade Universitária, Rio de Janeiro, 21941-850, RJ, Brazil}

\begin{abstract}
\textcolor{blue}{This is the preprint version of the published manuscript \url{https://doi.org/10.1016/j.cpc.2025.109842}. Please cite as:
Lange, R.; Magalhães, G.M.; Rocha, F.F.; Ribeiro Neto, H., 2025. An advanced fully-implicit solver for heterogeneous porous media based on foam-extend. Computer Physics Communications, 317, 109842.}

Multiphase flow in porous media is present in many engineering applications, including hydrogeology, oil recovery, and CO$_2$ sequestration. Accurate predictions of fluid behavior in these systems can improve process efficiency while mitigating environmental and health risks. 
Commercial simulators and open source software, such as the \texttt{porousMultiphaseFoam} repository based on the OpenFOAM framework, have been developed to model this type of problem. However, simulating heterogeneous porous media with heterogeneous porosity and permeability distributions poses significant numerical challenges. We introduce \texttt{coupledMatrixFoam}, an OpenFOAM-based solver designed for enhanced numerical stability and robustness. 
\texttt{coupledMatrixFoam} integrates the Eulerian multi-fluid formulation for phase fractions with Darcy’s law for porous media flow, applying a fully implicit, block-coupled solution for pressure and phase fractions. The solver is based on foam-extend 5.0, leveraging the latest \texttt{fvBlockMatrix} developments to improve computational efficiency. This approach enables a significant increase in time step sizes, particularly in cases involving capillary pressure effects and other complex physical interactions.
This work details the formulation, implementation and validation of \texttt{coupledMatrixFoam}, including comparisons with \texttt{porousMultiphaseFoam} that uses a segregated approach, to assess performance improvements. Additionally, a scalability analysis is conducted, demonstrating the solver’s ability for high-performance computing (HPC) applications, which are essential for large-scale, real-world simulations.\\

\end{abstract}

   
\end{frontmatter}


\section{Introduction}

Multiphase flows in porous media systems are related to a great variety of engineering applications such as oil production, underground aquifers, CO$_2$ sequestration, fuel cells, porous concrete, bed filters, and others \cite{bear1988dynamics}. Several computational models for simulating such types of flows have been proposed \cite{chen2006computational}, and open-source porous media simulators developed. Examples of open-source frameworks and libraries are DuMux \cite{Kochetal2020Dumux}, MRST \cite{krogstad2015mrst}, OpenGeoSys \cite{ogsRef}, PFlotran \cite{pflotran-paper}, Open Porous Media \cite{rasmussen2021open}, and MOOSE (PorousFlow module) \cite{Wilkins2020}. In the same way as the aforementioned initiatives, this paper presents an open repository that contains a set of computational codes for porous media flow simulations. The repository, called \texttt{porousMedia}, has been developed by WIKKI Brasil in the OpenFOAM framework \cite{jasak1996error, weller1998tensorial}.

Several open-source codes, such as \cite{beale2016open, maes2021geochemfoam, missios2023extending, soulaine2021porousmedia4foam, guibert2025micro, icardi2023computational}, have been developed within the OpenFOAM framework to simulate porous media applications, addressing various physical and numerical challenges. Considering those related to oil production, which is our main application of interest, we can mention the well-known \texttt{porousMultiphaseFoam} toolbox, introduced in \cite{horgue2015open} for the simulation of incompressible two-phase flows in porous media, and then extended to adaptive mesh refinement \cite{sugumar2020grid}, black-oil model \cite{fioroni2021openfoam}, hydrogeological ﬂows modeling \cite{horgue2022porousmultiphasefoam}, two-phase systems containing porous and solid-free regions \cite{carrillo2020multiphase}, and fracture models \cite{sangnimnuan2021development}. In this work, we present the \texttt{coupledMatrixFoam}, a new OpenFOAM solver for the simulation of fully coupled multiphase flow and transport in heterogeneous porous media, typical of oil production scenarios. The introduced solver is based on the Euler-Euler multi-fluid model for physical systems of $n$ phase fractions \cite{ishii2010thermo}, as well as the solver we presented in the preliminary work \cite{upstream_paper}. Our solvers differ from other approaches by modeling the porous medium as a stationary phase that composes the system.  Moreover, here we address a unique contribution in terms of computational efficiency by solving pressure and all phase fractions simultaneously throughout a fully implicit implementation that applies the block matrix framework of foam-extend version 5.0 \cite{uroic2019implicitly}. 

\texttt{coupledMatrixFoam} solver considers the Euler-Euler model for a system of phase fractions combined with Darcy’s law for flows through porous media \cite{muskat1981physical}, defining the rock system as a set of stationary phases. 
The code implementation is based on the very used \texttt{multiphaseEulerFoam} application \cite{weller2008new}, a choice that facilitates future extensions, including multiple species, heat and mass transfer, and other physical effects. In line with our previous works \cite{upstream_paper} and \cite{rocha2023openfoam}, the proposed solver has important functionalities related to porous media modeling and simulation, including support for heterogeneous permeability and porosity fields and specialized models for relative permeability and capillary pressure. Furthermore, the solution approach introduced here couples the phase fractions and pressure solution in order to achieve substantial improvements in computational efficiency for scenarios characterized by high complexity in terms of porosity, permeability, and capillary pressure effects.

For simulating multiphase flows there are generally sequential and coupled approaches. Sequential approaches, such as the Implicit Pressure Explicit Saturation (IMPES, \cite{coats2003impes}), simplify the numerical solution procedures by splitting operators, but presents severe time step limitations. On the other hand, coupled schemes are more complex as they solve the same system of equations simultaneously, but guarantee stability enabling higher advances in time \cite{monteagudo2007comparison}. Newton methods combined with preconditioner strategies are popular for solving multiphase flow equations implicitly in the context of porous media \cite{sun2013fully, yang2018scalable, luo2020fully}. In the current work, we use Taylor series expansions to linearize pressure and phase fractions dependencies, solving the resulting linearized system by a structure based on the latest development aimed at coupled solutions in foam-extend (\texttt{fvBlockMatrix}). Analogous linearization procedures have been used to solve velocity and stress tensor coupling for laminar viscoelastic flows in \cite{fernandes2019coupled,magalhaes2025enhancing}, and to treat implicit relation between phase fractions for multiphase flows in \cite{keser2019implicitly, keser2021eulerian}. Related works that use the block-coupled matrix structure of foam-extend include a transient implicit pressure-velocity coupling scheme for the Eulerian fluid model with any number of phases \cite{ferreira2019implementation}; a selective algebraic multigrid solver applied to the implicitly coupled steady-state incompressible pressure-velocity system \cite{uroic2018block}; and a coupled pressure based solution algorithm to model interface capturing problems \cite{kissling2010coupled}. Note that the works developed in OpenFOAM cited above do not deal with applications in porous media, which reinforces the importance of the cases to be studied using the \texttt{coupledMatrixFoam}. 

The fully implicit algorithm developed has the potential of improving code's robustness significantly when compared to segregated strategies. The effectiveness of the implemented methodology is demonstrated by notable improvements in the average time step, resulting in substantial reductions in computational costs. Additionally, a scalability analysis is conducted, demonstrating the solver’s suitability for high-performance computing (HPC) applications. Therefore, this work contributes to the advancement of computational efficiency in solving multiphase flows in porous media, particularly in scenarios with complex geometries and significant capillary pressure effects.

This paper provides a comprehensive description of the solver's formulation, implementation, and validation, with its main novelties summarized as follows:
\begin{itemize}
    \item Fully implicit coupling of phase fraction and pressure equations;
    \item Implicit treatment of Darcy and capillary pressure terms;
    \item Enhanced numerical stability compared to segregated approaches;
    \item Open-access repository with all validation and application cases.
\end{itemize}

The paper is divided into the following sections. The mathematical model is presented in Section \ref{sec:mathematical_model}, followed by the numerical formulation and implementation aspects in Section \ref{sec:numerical_formulation}. Detailed information about the computational code are provided in Section \ref{sec:computational_code}, numerical results in Section \ref{sec:results}, and conclusions in Section \ref{sec:conclusion}.

\section{Mathematical model}
\label{sec:mathematical_model}

\subsection{General equations}
\label{sub:general_eqs}

Applying the momentum conservation principle to each phase, we obtain Darcy's law, which describes fluid flow in porous media as follows:
\begin{equation}
\alpha_i \mathbf{U}_i = - \frac{k_{r,i}}{\mu_i} \mathbf{K} \cdot  ( \nabla p_i  - \rho_i \mathbf{g}) 
\label{eq:mom_Darcy}
\end{equation}
where $\mathbf{U}$ is the velocity, $\alpha$ is the phase fraction, $k_r$ the relative permeability, $\mathbf{K}$ the absolute permeability, $p$ the total pressure, $\rho$ the density, $\mu$ is the dynamic viscosity and $\mathbf{g}$ the gravity. The subscript $i$ indicates the phase. Considering a system of compressible phases, the mass balance equation for each phase is given as:
\begin{equation}
\frac{\partial (\alpha_i \rho_i)}{\partial t} +  \nabla \cdot \left( \alpha_i \rho_i \mathbf{U}_i \right) = q_i,
\label{eq:massBalance0}
\end{equation}
where $q_i$ is the source term related to the phase $i$.  Equation \eqref{eq:massBalance0} can be expanded and rewritten as
\begin{equation}
\frac{\partial \alpha_i}{\partial t}+\nabla \cdot\left(\alpha_i \mathbf{U}_i\right)
=
\frac{q_i}{\rho_i}-\frac{\alpha_i}{\rho_i} \frac{D_i \rho_i}{D t},
\label{eq:massBalance01}
\end{equation}
where the material derivative of $\rho$ accounts for the compressibility effects.

Substituting \equationautorefname \eqref{eq:mom_Darcy} in \equationautorefname \eqref{eq:massBalance01}, the new form of the phase continuity equation is given by:

\begin{equation}
\frac{\partial \alpha_i}{\partial t} -  \nabla \cdot \left(\frac{k_{r,i}}{\mu_i} \mathbf{K} \cdot  ( \nabla p_i  - \rho_i \mathbf{g}) \right) 
= 
\frac{q_i}{\rho_i}-\frac{\alpha_i}{\rho_i} \frac{D_i \rho_i}{D t}.
\label{eq:continuityDarcy}
\end{equation}
\equationautorefname \eqref{eq:continuityDarcy} is the final equation for the transport of each phase fraction. Note that the velocity fields were removed and it is no longer necessary to solve them.

It is important to highlight that in multiphase systems there exists the constraint 
\begin{equation}
    \sum_{p = 1}^{N_P} \alpha_p = 1,
    \label{eq:phaseConstraint1}
\end{equation}
where $N_P$ is the total number of phases in the multiphase system. Another relevant concept is the mixture velocity $\mathbf{U}_m$ which, for $n$ phases, is given by
\begin{equation}
\mathbf{U}_m = \sum_{i=1}^n \alpha_i \mathbf{U}_i.
\label{eq:Umix}
\end{equation}

The mass conservation of the compressible system by considering the mixture velocity can be defined as 
\begin{equation}
\nabla \cdot \left( \mathbf{U}_m \right) = 
\sum_{i=1}^{N_P} \frac{q_i}{\rho_i}-\sum_{i=1}^{N_P}\frac{\alpha_i}{\rho_i} \frac{D_i \rho_i}{D t}.
\label{eq:divUm}
\end{equation}

Substituting \equationautorefname \eqref{eq:mom_Darcy} in \equationautorefname \eqref{eq:Umix} and then, the resultant equation in \equationautorefname \eqref{eq:divUm}, we obtain:

\begin{equation}
- \nabla \cdot \left[ \sum_{i=1}^{N_P} \left( \frac{k_{r,i}}{\mu_i} \mathbf{K} \cdot  ( \nabla p_i  - \rho_i \mathbf{g}) \right) \right] 
= 
\sum_{i=1}^{N_P} \frac{q_i}{\rho_i}-\sum_{i=1}^{N_P}\frac{\alpha_i}{\rho_i} \frac{D_i \rho_i}{D t}.
\label{eq:divUM_darcy}
\end{equation}

Usually in OpenFOAM codes, \equationautorefname \eqref{eq:divUM_darcy} is used as the pressure equation of the system. In summary, the mathematical model for compressible multiphase flows in porous media is composed by the transport equations of phase fractions, \equationautorefname \eqref{eq:continuityDarcy}, the relationship between phase fractions, \equationautorefname \eqref{eq:phaseConstraint1}, and the pressure equation, \equationautorefname \eqref{eq:divUM_darcy}.

\subsection{Stationary phases}

Aiming to explore the advantages of the Euler-Euler methodology, the porous media is modeled as a stationary phase. Considering a null velocity for this type of phase, the continuity equation is given by:
\begin{equation}
    \dfrac{\partial \alpha_s}{\partial t} 
    = 
   \frac{q_s}{\rho_s}-\frac{\alpha_s}{\rho_s} \frac{D_s \rho_s}{D t},
    \label{eq:stationary}
\end{equation}
where the sub-index $s$ indicates a stationary phase and, as in the moving phases, the terms in the right side are used to account effects like compressibility, mass transfer, and other effects. 
It is possible to observe that \equationautorefname \eqref{eq:stationary} is a diagonal equation, with simple and cheap solution. 

Unlike traditional computational codes based on porosity ($\phi$), the solver \texttt{coupledMatrixFoam} uses phase fractions. The relation is given by
\begin{equation}
    \phi = \alpha_v,
\end{equation}
where $\alpha_v$ is the void volumetric fraction, calculated as
\begin{equation}
    \alpha_v = 1 - \sum_{k = 1}^{N_s} \alpha_k,
\end{equation}
considering $N_s$ the number of stationary phases in the system.

The consideration of stationary phases to model the porous media is strategically interesting as it is possible to use the foam-extend structure for the phase system both in these phases. 

\subsection{Effects of relative permeability}
\label{sub:krel_effects}

A comprehensive framework for multiphase flows, involving any number of phases, necessitates relative permeability models for each fluid-rock interface. The widely used Brooks and Corey (BC) model \cite{brooks1965hydraulic} relates the relative permeability of each phase to the phase fraction as follows:
\begin{equation}
k_{r,i} = k_{r,i (\text{max})} \left( \frac{\alpha_i}{\alpha_v} \right)^\eta,
\label{eq:kr_BandC}
\end{equation}
where $\eta$ is a power coefficient associated with the porous media properties, $\alpha_v$ is the void volumetric fraction, and $k_{r,i (\text{max})}$ is the maximal relative permeability. 

By substituting \equationautorefname~\eqref{eq:kr_BandC} into the pressure \equationautorefname~\eqref{eq:divUM_darcy}, one obtains:
\begin{equation}
- \nabla \cdot \left[\sum_{i=1}^{N_P} \left( \frac{k_{r,i(\text{max})}}{\mu_i} \left( \frac{\alpha_i}{\alpha_v} \right)^\eta \mathbf{K} \cdot (\nabla p_i - \rho_i \mathbf{g}) \right)\right] 
= 
\sum_{i=1}^{N_P} \frac{q_i}{\rho_i}-\sum_{i=1}^{N_P}\frac{\alpha_i}{\rho_i} \frac{D_i \rho_i}{D t}.
\label{eq:pressEq_kr}
\end{equation}
In a similar way, substituting \equationautorefname~\eqref{eq:kr_BandC} into the transport \equationautorefname~\eqref{eq:continuityDarcy}:
\begin{equation}
\frac{\partial \alpha_i}{\partial t} -  \nabla \cdot \left(\frac{k_{r,i(\text{max})}}{\mu_i} \left( \frac{\alpha_i}{\alpha_v} \right)^\eta \mathbf{K} \cdot  ( \nabla p_i  - \rho_i \mathbf{g}) \right) 
= 
\frac{q_i}{\rho_i}-\frac{\alpha_i}{\rho_i} \frac{D_i \rho_i}{D t}.
\label{eq:transpEq_kr}
\end{equation}

Equations \eqref{eq:pressEq_kr} and \eqref{eq:transpEq_kr} introduce a nonlinear term between pressure and phase fraction in the system of equations, requiring special treatment. Additionally, the \texttt{coupledMatrixFoam} framework supports the use of tabulated methodology, where users can provide a table with the relative permeability values as a function of saturation $S_i = \alpha_i/\alpha_v$.

\subsection{Capillary effects}

The discontinuity between two moving phases in a porous multiphase flow gives rise to an additional relationship between pressure fields known as capillary pressure $p_c$, given by:
\begin{equation}
p_{c,ki} = p_k - p_i,
\label{eq:capillary_pressure}
\end{equation}
where $k$ is a reference phase for the capillary pressure and $i$ represents any other phase in the system. Thus, the pressure of the reference phase $k$ and the capillary pressure of the pair $ki$ are used to define the pressure of any phase $i$. Replacing the pressure of a phase, $p_i$, in \equationautorefname \eqref{eq:pressEq_kr} and \equationautorefname \eqref{eq:transpEq_kr}, respectively, by the definition of capillary pressure in \equationautorefname \eqref{eq:capillary_pressure}, one can write:
\begin{multline}
- \nabla \cdot \left[ \sum_{i=1}^{N_P} \left(\frac{k_{r,i(\text{max})}}{\mu_i} \left( \frac{\alpha_i}{\alpha_v} \right)^\eta \mathbf{K} \cdot ( \nabla p_k -
\nabla p_{c,ki} - \rho_i \mathbf{g}) \right) \right] 
= \\
\sum_{i=1}^{N_P} \frac{q_i}{\rho_i}-\sum_{i=1}^{N_P}\frac{\alpha_i}{\rho_i} \frac{D_i \rho_i}{D t},
\label{eq:final_momentum}
\end{multline}
and
\begin{equation}
\frac{\partial \alpha_i}{\partial t} -  \nabla \cdot \left(\frac{k_{r,i(\text{max})}}{\mu_i} \left( \frac{\alpha_i}{\alpha_v} \right)^\eta \mathbf{K} \cdot  ( \nabla p_k -
\nabla p_{c,ki} - \rho_i \mathbf{g}) \right)
= 
\frac{q_i}{\rho_i}-\frac{\alpha_i}{\rho_i} \frac{D_i \rho_i}{D t}.
\label{eq:final_transpEq_kr_pc}
\end{equation}

It is worth noting that for the reference phase, capillary pressure satisfies $p_{c,kk} = 0$, and hence Equations \eqref{eq:pressEq_kr} and \eqref{eq:transpEq_kr} remain unchanged.

A useful approximation to the capillary pressure is to consider that it depends only on saturation, and hence
\begin{equation}
\nabla p_{c,ki} = \frac{\partial p_{c,ki}}{\partial S_{i}} \nabla S_{i}.
\label{eq:pcSFunc}
\end{equation}

Since the mathematical formulation of the solver developed is based on phase fractions, not in saturation, so, one more manipulation is necessary for \equationautorefname~\eqref{eq:pcSFunc}. Considering that the saturation of a phase $i$ is given by $S_i = \alpha_i/\alpha_v$, we can write
\begin{equation}
\nabla p_{c,ki} = \frac{\partial p_{c,ki}}{\partial S_{i}} \left(\frac{\nabla \alpha_{i}}{\alpha_v} - \frac{\alpha_{i} \nabla \alpha_{v}}{\alpha_{v}^{2}} \right). 
\label{eq:pcAlphaFunc}
\end{equation}
In this way, it is possible to add the capillarity effect without adding new variables to the system of equations. However, the capillarity models must be able to provide the derivative of capillary pressure as a function of saturation.

\subsection{Compressibility effects}

For the consideration of compressibility effects in the phases of the system, we assume that the density can be expressed in terms of the pressure through compressibility terms. A basic model is implemented for compressible flows computing the density variation according to a constant compressibility, such that: 

\begin{equation}
    \rho = \rho_0 + \psi \, p,
    \label{eq:compressibility}
\end{equation}
where $\rho_0$ is a reference density, $p$ is the pressure and  $\psi$ is the compressibility factor.

It is possible to apply the model both for fluid and rock phases, allowing the use of the developed code for classical cases and real problems in porous media.

\section{Numerical formulation and implementation}
\label{sec:numerical_formulation}

Numerical formulation and implementation aspects applied to treat different terms to be inserted into the block system resulting from the implicit coupling of pressure and phase fractions will be described in this section.

\subsection{Linearization for non-linear terms}
\label{sec:linearization}
In the coupled formulation, non-linear terms require a linearization to construct a linear system of equations to be operated by the solver. The Taylor Approximation (TA) method, as applied by Keser et al. \cite{keser2019implicitly}, was employed to linearize these terms.

To use Taylor expansions of functions in the coupled problem considered here, we suppose that the variables $\alpha_i$ and $p$ can be expressed, respectively, as $\alpha_i^n = \alpha_i^o + \Delta \alpha_i$ and $p^n = p^o + \Delta p$, where the superscript $n$ indicates the new approximate value (in the current time-step/iteration), $o$ indicates the old value (from the previous time-step/iteration), and $\Delta$ is a small correction. The first-order Taylor series expansion of the pressure and phase fraction unknowns can be expressed as
\begin{equation}
    f(\alpha_i^n,p^n) \approx f(\alpha_i^o,p^o) + \Delta \alpha_i \dfrac{\partial f(\alpha_i^o,p^o)}{\partial \alpha_i} + \Delta p \dfrac{\partial f(\alpha_i^o,p^o)}{\partial p}.
    \label{eq:taylorExp0}
\end{equation}

Considering that the goal is to obtain the value in the current time, the corrections could be expressed as $\Delta \alpha_i = \alpha_i^n - \alpha_i^o$ and $\Delta p = p^n - p^o$. Besides this consideration, the function $f$ can be written in the following form:
\begin{equation}
    f(\alpha_i^n,p^n) =  \alpha_i^n \, p^n \, \Upsilon(\alpha_i^o,p^o),
    \label{eq:f2Ups}
\end{equation}
where $\alpha_i^n \, p^n$ is the implicit part of $f$, and $\Upsilon(\alpha_i^o,p^o)$ represents the explicit part. 
Substituting the variations and \equationautorefname~\eqref{eq:f2Ups} in \equationautorefname~\eqref{eq:taylorExp0}, the Taylor expansion is given by:
\begin{multline}
    f(\alpha_i^n,p^n) \approx \alpha_i^o \, p^o \, \Upsilon(\alpha_i^o,p^o) 
    + (\alpha_i^n - \alpha_i^o) \dfrac{\partial (\alpha_i^o \, p^o \, \Upsilon(\alpha_i^o,p^o))}{\partial \alpha_i} \\
    + (p^n - p^o) \dfrac{\partial (\alpha_i^o \, p^o \, \Upsilon(\alpha_i^o,p^o))}{\partial p},
    \label{eq:taylorExp1}
\end{multline}
that leads to
\begin{equation}
    f(\alpha_i^n,p^n) \approx 
      \alpha_i^n p^o \, \Upsilon(\alpha_i^o,p^o)
    + \alpha_i^o \, p^n \, \Upsilon(\alpha_i^o,p^o)
    - \alpha_i^o \, p^o \, \Upsilon(\alpha_i^o,p^o).
    \label{eq:taylorExp2}
\end{equation}
Therefore, TA divides the nonlinear terms involving pressure and phase fractions into three components: explicit, implicit, and cross-coupling. 

\subsection{Handling coupled unknowns}

Since the solution algorithm proposed here employs a linear solver, the non-linear terms must be linearized. The application of the methodology described in subsection \ref{sec:linearization} for treating nonlinear terms will be detailed following.

The non-linear terms to be treated are present in the pressure \equationautorefname\eqref{eq:final_momentum} and transport  
\equationautorefname\eqref{eq:final_transpEq_kr_pc}. 
To facilitate the implementation process, it is introduced an auxiliary variable $\mathbf{T}_i$ for each phase $i$ of the system, which is defined as:
\begin{equation}
\mathbf{T}_i = \frac{k_{r,i(\text{max})}}{\mu_i\alpha_i} \left( \frac{\alpha_i}{\alpha_v} \right)^\eta \mathbf{K}.
\label{eq:TauxDef}
\end{equation}

As $q_i$ is a general source term, it is better to consider, for generality, the right side of \equationautorefname\eqref{eq:final_momentum} and \equationautorefname\eqref{eq:final_transpEq_kr_pc} as a function of pressure and phase fraction. For the next steps in the formulation, it will be called $\Gamma_i(p,\alpha)$:
\begin{equation}
    \Gamma_i(p,\alpha) = \frac{q_i}{\rho_i}-\frac{\alpha_i}{\rho_i} \frac{D_i \rho_i}{D t}.
\end{equation}

Note that by using $\mathbf{T}_i$ and $\Gamma_i(p,\alpha)$ in \equationautorefname\eqref{eq:final_momentum} and \equationautorefname\eqref{eq:final_transpEq_kr_pc}, we obtain, respectively:
\begin{equation}
- \nabla \cdot \left[ \sum_{i=1}^{N_P} \left(\alpha_i \mathbf{T}_i \cdot ( \nabla p_k -
\nabla p_{c,ki} - \rho_i \mathbf{g}) \right) \right] 
= 
\sum_{i=1}^{N_P} \Gamma_i(p,\alpha_i),
\label{eq:final_momentum_Taux}
\end{equation}
and
\begin{equation}\frac{\partial \alpha_i}{\partial t} -  \nabla \cdot \left(\alpha_i \mathbf{T}_i \cdot  ( \nabla p_k -
\nabla p_{c,ki} - \rho_i \mathbf{g}) \right)
= 
\Gamma_i(p,\alpha_i).
\label{eq:final_transpEq_kr_pc_Taux}
\end{equation}
Therefore, the non-linear terms to be treated have been rewritten in the form
\begin{equation}
\alpha_i \mathbf{T}_i \cdot (\nabla p_k - \nabla p_{c,ki} - \rho_i \mathbf{g}).
    \label{eq:final_non-linear_terms}
\end{equation}

Next, we explain the different aspects to be considered for each nonlinear term of \equationautorefname\eqref{eq:final_non-linear_terms}.

\subsubsection{Pressure term}

The non-linear term involving phase fraction $\alpha_i$ and reference phase pressure, here denoted by $p$ for simplicity, can be handled using TA. To linearize such term, we assume that $\mathbf{T}_i$ can be calculated explicitly using the phase fraction approximation from the previous time-step/iteration (denoted by the superscript $o$). Thus, the non-linear term between phase fraction and pressure at the current time-step/iteration (denoted by the superscript $n$) is given by:
\begin{equation}
\left(\alpha_i \,\mathbf{T}_i \cdot \nabla p\right)^n \approx \mathbf{T}_i^o \cdot (\alpha_i^{n} \nabla p^o + \alpha_i^{o} \nabla p^n - \alpha_i^{o} \nabla p^o).
\label{eq:tag20}
\end{equation}

It is important to emphasize that the relative permeability model generally is highly dependent on phase fraction values, and hence, the implicit treatment for $\alpha_i$ in the product $\alpha_i \mathbf{T}_i$ is crucial to ensure improvements to the accuracy and stability of the system.

\subsubsection{Gravity term}

In phase fraction and pressure equations, there is also consideration of gravitational effects on the flow, modeled by the term $\alpha_i\,\rho_i\, \mathbf{T}_i\cdot \mathbf{g}$ in \equationautorefname\eqref{eq:final_non-linear_terms}. 
The simplest way to handle this term would be to add its effect explicitly. However, using an explicit approach to treat key terms in the system can lead to significant reductions in the time step, reducing the computational efficiency. An alternative way that we use to linearize it, which depends implicitly on the phase fraction, is given by:
\begin{equation}
\left(\alpha_i \,\rho_i\,\mathbf{T}_i \cdot \mathbf{g} \right)^n\approx \alpha_i^n \,\rho_i^o\,\mathbf{T}_i^o \cdot \mathbf{g}.
\label{eq:gravLin}
\end{equation}

Thus, the phase fraction is evaluated at the current time ($\alpha_i^n$), which avoids a fully explicit treatment of the gravity-related term, resulting in less impact on the stability and robustness of the implemented code.

\subsubsection{Capillary pressure term}

The non-linear term involving phase fraction and capillary pressure in \equationautorefname\eqref{eq:final_non-linear_terms} is $\alpha_i \mathbf{T}_i \cdot \nabla p_{c,ki}$. Substituting \equationautorefname \eqref{eq:pcAlphaFunc} into the referred term and rearranging it, the capillary pressure contribution can be written as:
\begin{equation}
\alpha_i \,\frac{\partial p_{c,i}}{\partial S_{i}}\,\mathbf{T}_i \cdot\left(\frac{\nabla \alpha_{i}}{\alpha_v} - \frac{\alpha_{i} \nabla \alpha_{v}}{\alpha_{v}^{2}} \right),
\label{eq:termPcTaux}
\end{equation}
where the sub-index $k$ related to the reference phase has been omitted for the sake of simplicity.

Equation \eqref{eq:termPcTaux} can be separated into two terms, and both need to be treated through linearizations. For the first term, one possibility is to treat the phase fraction multiplying the partial derivative of capillary pressure with respect to saturation as information obtained from the previous iteration ($\alpha_i^o$):
\begin{equation}
\left(\alpha_i \,\frac{\partial p_{c,i}}{\partial S_{i}}\,\mathbf{T}_i \cdot\frac{\nabla \alpha_{i}}{\alpha_v}\right)^n \approx \alpha_i^o \,\left(\frac{\partial p_{c,i}}{\partial S_{i}}\right)^o \,\mathbf{T}_i^o \cdot\frac{\nabla \alpha_{i}^n}{\alpha_v^o}.
\label{eq:explicitTerm1}
\end{equation}
The second alternative is to apply the TA, taking the phase fraction and the gradient of phase $i$ implicitly. Following this approach, the first term can be linearized as:
\begin{equation}
\left(\alpha_i\, \frac{\partial p_{c,i}}{\partial S_{i}}\, \mathbf{T}_i \cdot\frac{\nabla \alpha_{i}}{\alpha_v}\right)^n 
\approx 
\frac{1}{\alpha_v^o}\,\left(\frac{\partial p_{c,i}}{\partial S_{i}}\right)^o \, \mathbf{T}_i^o \cdot \left(
\alpha_i^o 
\nabla \alpha_{i}^n
+ \alpha_i^n 
\nabla \alpha_{i}^o
\
-\alpha_i^o 
\nabla \alpha_{i}^o
\right),
\label{eq:implicityTerm1}
\end{equation}
where the presence of three parcels: coupling, implicit, and explicit is observed when applying the methodology.

During the development of the code, both alternatives were compared in different scenarios and no difference was observed in terms of stability, robustness, or accuracy. So, considering that the first linearization is cheaper, it was the option implemented in the final version of the code. It is important to highlight that the second treatment could be beneficial for some future implementations in the code, so, the possibility to use the second treatment is open.

Finally, in the linearization of the second term in \equationautorefname \eqref{eq:termPcTaux}, only one of the phase fractions $\alpha_i$ will be considered at the current time, resulting in:
\begin{equation}
\left(-\alpha_i \alpha_i\,  \frac{\partial p_{c,i}}{\partial S_{i}} \,\mathbf{T}_i \cdot\frac{\nabla \alpha_{v}}{\alpha^2_{v}} \right)^n
\approx -\alpha_i^n \alpha_i^o \, \left(\frac{\partial p_{c,i}}{\partial S_{i}}\right)^o\,\mathbf{T}_i^o \cdot \left(\frac{\nabla \alpha_{v}}{\alpha_{v}^2} \right)^o.
\label{eq:linTerm2Pc}
\end{equation}

We remark that the void fraction $\alpha_v$ does not change throughout the entire simulation when effects such as compressibility in the porous medium are not considered. In such cases, the evaluation of the void fraction can always use the value from the previous iteration or time step. The same reasoning has been applied to all other terms where this parameter appears.

\subsubsection{Other effects}

The function $\Gamma_i(p,\alpha)$ is used to take account of effects related to fluid compressibility, mass transference, and other effects that generally appear as source terms in common explicit formulations \cite{upstream_paper}. An important note regarding $\Gamma_i(p,\alpha)$ is that, despite writing this right of the equal sign in the equations for pressure and phase fraction, it is not treated explicitly. The TA (section~\ref{sec:linearization}) is applied, obtaining explicit, implicit and cross-coupling parcels:
\begin{equation}
    \Gamma_i(p,\alpha) \approx \Gamma_i(p^n,\alpha^o) - \Gamma_i(p^o,\alpha^o) + \Gamma_i(p^o,\alpha^n).
\end{equation}

\subsection{The final system of equations}

With all terms properly linearized, it is possible to write the final system of equations, which will later be assembled into a block matrix. The final format for each equation of the system is presented below.

\subsubsection{Reference moving phase}
In order to ensure the constraint \equationautorefname\eqref{eq:phaseConstraint1}, a moving phase $\alpha_R$ is chosen as a reference phase and obtained by the relationship:  
\begin{equation}
    \alpha_R = 1 - \sum_{p = 1}^{N_p-1} \alpha_p,
    \label{eq:phaseConstraint}
\end{equation}
instead of being solved by the transport equation. It is also possible to solve all phases within the block system without setting this constraint, but we will show in the numerical experiments that the performance of the first case is significantly better.

\subsubsection{Other moving phases}
The final transport equation of each other moving phase to be considered inside the block-system is given by:
\begin{multline}
\frac{\partial \alpha_i}{\partial t} 
-\nabla\cdot\Big[ \mathbf{T}_i^o \cdot ( \alpha_i^{n} \nabla p^o + \alpha_i^{o}  \nabla p^n - \alpha_i^{o} \nabla p^o)
-\frac{1}{\alpha_v^o} \left(\frac{\partial p_{c,i}}{\partial S_{i}}\right)^o  \,\mathbf{T}_i^o \cdot \left(
\alpha_i^o \nabla \alpha_{i}^n 
\right) \\
+\alpha_i^n \alpha_i^o \left(\frac{\partial p_{c,i}}{\partial S_{i}}\right)^o\mathbf{T}_i^o \cdot \left(\frac{\nabla \alpha_{v}}{\alpha_{v}^{2}} \right)^o
- \alpha_i^n \rho_i^o\mathbf{T}_i^o \cdot \mathbf{g}\Big]
\\
 - \Gamma_i(p^o,\alpha^n) + \Gamma_i(p^o,\alpha^o) - \Gamma_i(p^n,\alpha^o)
= 0.
\label{eq:finalAlpha}
\end{multline}

\subsubsection{Stationary phases}
Since we consider the porous medium modeled as a stationary phase, the respective continuity equations must be approximated. Here, such equations are given by  
\begin{equation}
    \dfrac{\partial \alpha_s}{\partial t} = 
    0,
    \label{eq:stationary_final}
\end{equation}
and, hence, it is not necessary to be solved. 

For this first version of the solver, the effects of compressibility were only included for the moving phases, as for the multiphase problems in porous media, the incompressibility of the fluids will lead to results far from reality. The scenario for dilatation (rock compressibility) is quite different, as it is possible to assume the hypothesis of incompressibility in a range of cases with less impact in the accuracy \cite{chen2006computational}.

We intend to add the possibility of varying porous media
properties during the simulations in forthcoming works, which will include a non-zero source term to
\equationautorefname\eqref{eq:stationary_final}.

\subsubsection{Pressure equation}
The pressure equation, \equationautorefname\eqref{eq:final_momentum_Taux}, needs to be adapted to satisfy the closure \equationautorefname\eqref{eq:phaseConstraint} when a reference phase is defined. Therefore, for a physical system containing $N_P$ phases, the pressure equation considering only the first term (i.e., only the pressure gradient contribution), without the additional effects $\Gamma$, is given by:
\begin{equation}
-\nabla \cdot \Big[\sum_{\substack{i=1 \\ i \neq R}}^{N_P}
\Big(
(\mathbf{T}_i^o - \mathbf{T}_R^o ) \cdot ( \alpha_i^{n} \nabla p^o + \alpha_i^{o}  \nabla p^n - \alpha_i^{o} \nabla p^o)\Big)
+ \alpha_v^{o} \mathbf{T}_R^o \nabla p^n \Big]
= 
0.
\label{eq:press_final_sys}
\end{equation}

The gravity contribution, in turn, can be written as:
\begin{equation}
\nabla \cdot \Big[\sum_{\substack{i=1 \\ i \neq R}}^{N_P}
\Big(
\alpha_i^n (\rho_i^o\,\mathbf{T}_i^o - \rho_R^o\,\mathbf{T}_R^o) \cdot \mathbf{g}
\Big)
+ \alpha_v^{o} \rho_R^o\,\mathbf{T}_R^o \mathbf{g} \Big].
\label{eq:gravity_final_sys}
\end{equation}
Including gravity effects measured by \equationautorefname\eqref{eq:gravity_final_sys} into the pressure  \equationautorefname\eqref{eq:press_final_sys}, one obtains:
\begin{multline}
-\nabla \cdot \Big[\sum_{\substack{i=1 \\ i \neq R}}^{N_P}
\Big(
(\mathbf{T}_i^o - \mathbf{T}_R^o ) \cdot ( \alpha_i^{n} \nabla p^o + \alpha_i^{o}  \nabla p^n - \alpha_i^{o} \nabla p^o)
- \alpha_i^n (\rho_i^o\mathbf{T}_i^o - \rho_R^o\mathbf{T}_R^o) \cdot \mathbf{g}
\Big) \\
+ \alpha_v^{o} \mathbf{T}_R^o \nabla p^n 
- \alpha_v^{o} \rho_R^o\mathbf{T}_R^o \mathbf{g} \Big]
 = 0.
\label{eq:general_DandG}
\end{multline}

Finally, the capillary pressure contribution is determined by:
\begin{multline}
\nabla \cdot \Big[\sum_{\substack{i=1 \\ i \neq R}}^{N_P}
\left(\frac{1}{\alpha_v^o}
\Big(\dfrac{\partial p_{c,i}}{\partial S_i}\Big)^o \, \mathbf{T}_i^o \cdot
\left( \alpha_i^o\nabla\alpha_i^n \right)
- 
\alpha_i^n \alpha_i^o \Big(\frac{\partial p_{c,i}}{\partial S_i}\Big)^o\,\mathbf{T}_i^o \cdot \Big(\frac{\nabla \alpha_{v}}{\alpha_{v}^{2}}
\Big)^o\right)
\\
- \frac{1}{\alpha_v^o}\Big(\dfrac{\partial p_{c,R}}{\partial S_R}\Big)^o\, \alpha_R^o \,\mathbf{T}_R^o \cdot \left( 
 \sum_{\substack{i=1 \\ i \neq R}}^{N_P} \nabla\alpha_i^n \right) \\
+ \alpha_R^o \Big(\frac{\partial p_{c,R}}{\partial S_R}\Big)^o\,\mathbf{T}_R^o \cdot \Big(\frac{\nabla \alpha_{v}}{\alpha_{v}^{2}}
\Big)^o  \left( \sum_{\substack{i=1 \\ i \neq R}}^{N_P} \alpha_i^n \right) \Big]
.
\label{eq:general_pc}
\end{multline}

Using the general form of \equationautorefname~\eqref{eq:general_DandG} and \equationautorefname~\eqref{eq:general_pc}, the final shape of pressure equation, also accounting the additional effects $\Gamma$, where the compressibility for the moving phases is included, is given by:
\begin{multline}
-\nabla \cdot \Big[\sum_{\substack{i=1 \\ i \neq R}}^{N_P}
\Big(
(\mathbf{T}_i^o - \mathbf{T}_R^o ) \cdot ( \alpha_i^{n} \nabla p^o + \alpha_i^{o}  \nabla p^n - \alpha_i^{o} \nabla p^o)
- \alpha_i^n (\rho_i^o\mathbf{T}_i^o - \rho_R^o\mathbf{T}_R^o) \cdot \mathbf{g}
\Big)  \\
+ \alpha_v^{o} \mathbf{T}_R^o \nabla p^n 
- \alpha_v^{o} \rho_R^o\mathbf{T}_R^o \mathbf{g} \\
-\sum_{\substack{i=1 \\ i \neq R}}^{N_P}
\left(\frac{1}{\alpha_v^o}
\Big(\dfrac{\partial p_{c,i}}{\partial S_i}\Big)^o \, \mathbf{T}_i^o \cdot
\left(
\alpha_i^o\nabla\alpha_i^n 
\right)
- 
\alpha_i^n \alpha_i^o \Big(\frac{\partial p_{c,i}}{\partial S_i}\Big)^o\,\mathbf{T}_i^o \cdot \Big(\frac{\nabla \alpha_{v}}{\alpha_{v}^{2}}
\Big)^o\right)
\\
- \frac{1}{\alpha_v^o}\Big(\dfrac{\partial p_{c,R}}{\partial S_R}\Big)^o\, \alpha_R^o \,\mathbf{T}_R^o \cdot \left( 
 \sum_{\substack{i=1 \\ i \neq R}}^{N_P} \nabla\alpha_i^n \right)
+ \alpha_R^o \Big(\frac{\partial p_{c,R}}{\partial S_R}\Big)^o\,\mathbf{T}_R^o \cdot \Big(\frac{\nabla \alpha_{v}}{\alpha_{v}^{2}}
\Big)^o  \left( \sum_{\substack{i=1 \\ i \neq R}}^{N_P} \alpha_i^n \right) \Big] 
\\
 - \sum_{\substack{i=1 \\ i \neq R}}^{N_P}\Gamma_i(p^o,\alpha^n) - \sum_{\substack{i=1 \\ i \neq R}}^{N_P}\Gamma_i(p^n,\alpha^o) + \sum_{\substack{i=1 \\ i \neq R}}^{N_P} \Gamma_i(p^o,\alpha^o) 
    = 0.
\label{eq:finalPress_gen}
\end{multline}

It is important to highlight that the code can be used with or without a reference phase. If a reference phase is not defined, all the terms with subscript $R$ are omitted in \equationautorefname~\eqref{eq:finalPress_gen}.

\subsection{Special numerical treatments}

As the current work is based on a complex mathematical formulation with implicit coupling between equations, it is crucial to carefully select and combine appropriate numerical methods. The solver must simultaneously resolve multiple physical mechanisms, including capillary pressure effects, gravitational force, and pressure gradients. Therefore, the discretization schemes and interpolation treatments used in each term directly influence the overall stability and accuracy of the solution.

During the development of the solver, a series of tests were conducted to understand the impact of various numerical treatments on stability, especially in challenging scenarios. One of the key objectives was to exploit the benefits of the Darcy equation in the proposed formulation. Since Darcy's law can be treated implicitly in the momentum balance, the corresponding term appears in the diagonal of the momentum matrix. If not handled correctly, this implicit term can lead to numerical inconsistencies, particularly regarding the choice of interpolation scheme.

\subsubsection{Face interpolation schemes}
\label{sec:face_interpolation}

In cell-centered finite volume approaches, certain variables must be interpolated from cell centers to cell faces to maintain consistency and accuracy in numerical calculations. OpenFOAM provides several interpolation methods, including linear, harmonic, and upwind schemes, each suited for specific variables and physical scenarios.

A well-known example in porous media flows is the use of harmonic interpolation for permeability. This choice is grounded in physical and mathematical principles. For instance, a simple one-dimensional test with two distinct permeability values demonstrates that harmonic interpolation precisely reproduces the analytical solution dictated by Darcy’s law \cite{chen2006computational}. This result highlights the importance of selecting the appropriate interpolation scheme to honor the underlying physics of the problem.

On the other hand, linear interpolation is commonly used in OpenFOAM for various other quantities. For example, standard tutorial cases often apply linear interpolation to the inverse of the diagonal in the momentum system. However, when the Darcy term is implicitly embedded in the momentum matrix, linear interpolation can lead to numerical inconsistencies. The velocity field in porous media is highly sensitive to the permeability distribution, and an inappropriate interpolation scheme may compromise solution quality. In contrast, harmonic interpolation more accurately captures the flow resistance in media with strong heterogeneities or sharp permeability variations, ensuring better fidelity to the physical system.

In this context, it is worth emphasizing that $\mathbf{T}_i$ (as defined in Eq. \eqref{eq:TauxDef}) is inherently calculated at cell faces. This is because it relies on other face-centered properties, such as absolute permeability, which is itself computed using harmonic interpolation. This approach ensures that the relevant transport properties are accurately represented at the interfaces where fluxes are evaluated.

\subsubsection{Interpolation and divergence schemes}
\label{sec:TA_treatments}

Interpolation plays a critical role not only for physical properties but also in the treatment of flux calculations and divergence terms, which require face-centered values in OpenFOAM. For instance, phase fractions must also be interpolated to cell faces to maintain consistency, as their fluxes directly influence divergence terms. In the solver, each driving force (pressure gradient, gravity, and capillarity) is associated with a separate flux. This separation allows the use of distinct discretization schemes tailored to each term.

A key finding in the solver is that maintaining the closure of volume fractions requires the use of the same numerical flux across all phase fraction interpolations when divergence terms are discretized. For example, when the pressure is linearized with respect to the phase fractions, the interpolated values of all phase fractions and the pressure gradient must be consistently aligned to ensure numerical compatibility and accuracy. In this work, the proposed fully implicit formulation adopts the first-order upwind method.

Although fully implicit first-order approximations are the most widely used in general reservoir simulation \cite{abushaikha2017fully}, the generic implementation of interpolated fields in \texttt{coupledMatrixFoam} allows for the use of all numerical schemes available in OpenFOAM. For example, high-order total variation diminishing (TVD) schemes can be employed in specific cases where accurate resolution of discontinuities, such as shocks and phase interfaces, is crucial \cite{KOLGAN20112384, van1979towards, van2011historical}. Due to its versatility and extensibility, the developed solver is suitable for a variety of reservoir applications. The use of high-order methods and their impact on solution accuracy will be investigated in future work.

\subsubsection{The momentum reconstruction}
\label{sec:momentum_reconstruction}

In finite volume methods, the pressure field is typically stored at cell centers, while flux calculations require velocity (or momentum) values at the cell faces. These face-centered values are often obtained either by solving a momentum equation or by interpolating cell-centered properties. However, in the case of porous media modeled by Darcy’s law, an alternative approach leverages the interpolation of permeability to the faces using harmonic interpolation. This method provides an analytical solution for face-centered momentum. Other properties, such as density and viscosity, are interpolated linearly, which is generally an effective and reliable choice for these quantities.

After solving for pressure and phase fractions, the velocity field is obtained using OpenFOAM's $reconstruct$ function. This function calculates the cell-centered velocity by reconstructing it from the fluxes at the cell faces. The key principle here is that momentum conservation is ensured at the faces, while the cell-centered velocity values are derived from these fluxes rather than being directly modeled by the momentum equation.

While this approach may introduce small deviations in momentum conservation at the cell centers, these deviations do not affect the algorithm's accuracy, since the cell-centered velocity is not directly used in subsequent calculations. Instead, the flux values at the faces are the primary concern, as they play a critical role in ensuring that the solver produces high-quality and consistent results.

\subsection{Block-system building}
\label{sec:blockBuild}
To understand how the block system is built, the terms of the phase-fraction equation (\equationautorefname\eqref{eq:finalAlpha}) and the pressure equation (\equationautorefname\eqref{eq:finalPress_gen}) are highlighted in different colors. The colors are used to establish a correspondence with the system in \equationautorefname\eqref{eq:parallelSys}.

\begin{multline}
    \highlight[yellow!15]{\frac{\partial \alpha_i}{\partial t}
    -\nabla\cdot\Big[ \mathbf{T}_i^o \cdot ( \alpha_i^{n} \nabla p^o)\Big]}
    \highlight[green!15]{-\nabla\cdot\Big[ \mathbf{T}_i^o \cdot ( \alpha_i^{o}  \nabla p^n)\Big]}
    +\nabla\cdot\Big[ \mathbf{T}_i^o \cdot ( \alpha_i^{o} \nabla p^o)\Big]\\
    \highlight[yellow!15]{+\nabla\cdot\Big[ \frac{1}{\alpha_v^o} \left(\frac{\partial p_{c,i}}{\partial S_{i}}\right)^o  \,\mathbf{T}_i^o \cdot \left(
    \alpha_i^o \nabla \alpha_{i}^n 
    \right) }
    \\
    \highlight[yellow!15]{-\alpha_i^n \alpha_i^o \left(\frac{\partial p_{c,i}}{\partial S_{i}}\right)^o\mathbf{T}_i^o \cdot \left(\frac{\nabla \alpha_{v}}{\alpha_{v}^{2}} \right)^o
    + \alpha_i^n \rho_i^o\mathbf{T}_i^o \cdot \mathbf{g}\Big]
     - \Gamma_i(p^o,\alpha^n)} 
    \\ \highlight[green!15]{- \Gamma_i(p^n,\alpha^o)} + \Gamma_i(p^o,\alpha^o)
    = 
    0.
    \label{eq:finalAlphaCol}
\end{multline}

\begin{multline}
    \highlight[blue!35]{\nabla \cdot \Big[\sum_{\substack{i=1 \\ i \neq R}}^{N_P}
    \Big(
    (\mathbf{T}_i^o - \mathbf{T}_R^o ) \cdot ( \alpha_i^{n} \nabla p^o)
    \Big) \Big]}
    \highlight[red!15]{+ \nabla \cdot \Big[\sum_{\substack{i=1 \\ i \neq R}}^{N_P}
    \Big(
    (\mathbf{T}_i^o - \mathbf{T}_R^o ) \cdot (\alpha_i^{o}  \nabla p^n)
    \Big) \Big]}
    \\
    -\nabla \cdot \Big[\sum_{\substack{i=1 \\ i \neq R}}^{N_P}
    \Big(
    (\mathbf{T}_i^o - \mathbf{T}_R^o ) \cdot (\alpha_i^{o} \nabla p^o)
    \Big) \Big] 
    \highlight[blue!35]{- \nabla \cdot \Big[\sum_{\substack{i=1 \\ i \neq R}}^{N_P}
    \Big(
    \alpha_i^n (\rho_i^o\mathbf{T}_i^o - \rho_R^o\mathbf{T}_R^o) \cdot \mathbf{g}
    \Big) \Big]}
    \\
    \highlight[red!15]{+ \nabla \cdot \Big[\alpha_v^{o} \mathbf{T}_R^o \nabla p^n \Big]}
    - \nabla \cdot \Big[\alpha_v^{o} \rho_R^o\mathbf{T}_R^o \mathbf{g} \Big]
    \\
    \highlight[blue!35]{-\nabla \cdot \Big[\sum_{\substack{i=1 \\ i \neq R}}^{N_P}
    \left(\frac{1}{\alpha_v^o}
    \Big(\dfrac{\partial p_{c,i}}{\partial S_i}\Big)^o \, \mathbf{T}_i^o \cdot
    \left(
    \alpha_i^o\nabla\alpha_i^n 
    \right)
    - 
    \alpha_i^n \alpha_i^o \Big(\frac{\partial p_{c,i}}{\partial S_i}\Big)^o\,\mathbf{T}_i^o \cdot \Big(\frac{\nabla \alpha_{v}}{\alpha_{v}^{2}}
    \Big)^o\right)\Big]}
    \\
    \highlight[blue!35]{- \nabla \cdot \Big[\frac{1}{\alpha_v^o}\Big(\dfrac{\partial p_{c,R}}{\partial S_R}\Big)^o\, \alpha_R^o \,\mathbf{T}_R^o \cdot \left( 
     \sum_{\substack{i=1 \\ i \neq R}}^{N_P} \nabla\alpha_i^n \right)}
    \\
    \highlight[blue!35]{+ \alpha_R^o \Big(\frac{\partial p_{c,R}}{\partial S_R}\Big)^o\,\mathbf{T}_R^o \cdot \Big(\frac{\nabla \alpha_{v}}{\alpha_{v}^{2}}
    \Big)^o  \left( \sum_{\substack{i=1 \\ i \neq R}}^{N_P} \alpha_i^n \right) \Big] - \sum_{\substack{i=1 \\ i \neq R}}^{N_P}\Gamma_i(p^o,\alpha^n)}
    \\ \highlight[red!15]{- \sum_{\substack{i=1 \\ i \neq R}}^{N_P}\Gamma_i(p^n,\alpha^o)} + \sum_{\substack{i=1 \\ i \neq R}}^{N_P} \Gamma_i(p^o,\alpha^o) 
    = 
    0.
\label{eq:finalPress_genCol}
\end{multline}

Considering a system with the porous media modeled by $\alpha_s$ with a two-phase flow where the second phase is adopted as the reference phase  ($\alpha_R = \alpha_2$), the block-system in each cell is given by:

\begin{equation}
    \begin{bNiceMatrix}[margin]
        \Block[fill=red!15]{1-1}{}
        A_{p,p} & 
        \Block[fill=blue!35]{1-1}{}
        A_{p,\alpha_1} \\
         \Block[fill=green!15]{1-1}{}
         A_{\alpha_1,p} &
         \Block[fill=yellow!15]{1-1}{}
         A_{\alpha_1,\alpha_1}
    \end{bNiceMatrix}.
    \begin{bNiceMatrix}[margin]
        p \\
        \alpha_1
    \end{bNiceMatrix}
    =
    \begin{bNiceMatrix}[margin]
        b_{p} \\
         b_{\alpha_1}
    \end{bNiceMatrix}.
    \label{eq:parallelSys}
\end{equation}  

The colors are used to characterize the pressure implicit terms (red), the phase fraction implicit terms (yellow), the coupling of $\alpha_1$ in pressure (blue), and the coupling of pressure in the $\alpha_1$ equation (green). Then, the colored terms of the equations are inserted in the portion of the system with the same color. The white terms are explicitly treated, meaning that these are inserted in $b_p$ for the \equationautorefname~\eqref{eq:finalPress_genCol} or in $b_{\alpha_1}$ for the \equationautorefname~\eqref{eq:finalAlphaCol}.

\section{Computational code}
\label{sec:computational_code}

\subsection{The flexible structure to build the block-system}

One of the main characteristics of the \texttt{coupledMatrixFoam} is that the amount of coupled variables is not fixed. In other available block-coupled solvers in foam-extend the quantity of coupled variables doesn't change (there are four variables when using pressure-velocity coupling). Here, the main goal is the flexibility to consider different systems with different numbers of moving phases inside the block system. For this purpose, it was necessary to implement a new structure for the control operations involving the block matrices.

All the operations implemented in this context are in the folder \texttt{blockSystems} that have the sub-folders \texttt{block\_}, for the specific sizes of the system and \texttt{blockSystem}. In the current version, it is possible to consider block matrices with sizes from 2 to 5, remembering that the stationary phases are not solved and one moving phase can be obtained from the constraint (\equationautorefname \eqref{eq:phaseConstraint}). It means that in the current version, it is possible to consider multiphase systems from two to five phases (considering one as the reference phase for the last) as the first position in the block system is always used by pressure.

The counting of the number of equations in the block matrix is done in the \texttt{createFields.H} where the block size is defined and the index for each variable in the block system is stored (see code listing \ref{lst:pAsBlockCreate}). It is important to observe that the size of system just increase if the phase is moving and it is not the reference phase.

\begin{lstlisting}[
    style=customcpp, 
    caption=Building the pAs block matrix in \textit{createFields.H}, 
    label={lst:pAsBlockCreate}
]
label Ai = 0;
forAll(phases, phasei)
{
    phaseModel& phase = phases[phasei];

    if (!phase.stationary() && !phase.isReference())
    {
        Ai++;
        phase.blockIndex().append(Ai);
    }
}
const label blockSize = ++Ai;

// set size of phi0s
fluid.phi0s().setSize(blockSize);

// Block vector field for pressure (first entry) and phases fractions 
// (others entries).   
autoPtr<blockSystem> pAsBlock(blockSystem::New(blockSize, pAsDict, mesh)); 
\end{lstlisting}

Within the \texttt{pAsBlock} created, the positions of the block system are initialized using the same index order that will be applied along all the code: pressure in the first position (line 2 in code listing \ref{lst:pAsBlockInit}) and \texttt{Ai} defined getting the block index for each phase (lines 16 and 17 in code listing \ref{lst:pAsBlockInit}).

\begin{lstlisting}[
    style=customcpp, 
    caption=Initialization of pAs block matrix in \textit{createFields.H}, 
    label={lst:pAsBlockInit}
]
// Initialise p in pAs
pAsBlock->blockAdd(0, p.internalField());

forAll(phases, phasei)
{
    phaseModel& phase = phases[phasei];
    const volScalarField& alpha = phase;

    phase.alphafRef() = fvc::interpolate(alpha);

    if (phase.stationary() || phase.isReference()) continue;

    alphaTable.add(alpha);

    // Initialise alpha in pAs
    label Ai = phase.blockIndex()[0];
    pAsBlock->blockAdd(Ai, alpha.internalField());
}
\end{lstlisting}

After building and initializing the block system, the algorithm of the solution began. It will be explained in section \ref{sec:impl_solver}.

\subsection{General overview on \texttt{coupledMatrixFoam}}
\label{sec:impl_solver}
\begin{figure}[!h]
	\tikzstyle{block} = [rectangle, draw,
	text width=23em, text centered, rounded corners, minimum height=2em]
	\tikzstyle{block1} = [rectangle, draw,
	text width=13em, text centered, rounded corners, minimum height=2em]
	\tikzstyle{block2} = [rectangle, draw, fill=gray!20, 
	text width=4em, text centered, rounded corners, minimum height=2em]
    \tikzstyle{noborder} = [rectangle, text width=2em, text centered, minimum height=2em]
	\tikzstyle{line} = [draw, -latex']
	\tikzstyle{line2} = [draw, -]

	\begin{center}	
		\begin{tikzpicture}[node distance = 0.4cm, auto]
		\node [block1] (PC0) {Begin of external iteration};
		\node [block, below of=PC0, node distance=1.2cm] (PC1) {Initialize the pAs block system (matrix, source and reference to pAs)};
		\node [block, below of=PC1, node distance=1.4cm] (PC2) {Assemble and insert equations (phase fractions and pressure)};
		\node [block, below of=PC2, node distance=1.3cm] (PC3) {Solve the block matrix};
		\node [block, below of=PC3, node distance=1.3cm] (PC4) {Retrieve the solution to a separated field and update boundary conditions};
		\node [block, below of=PC4, node distance=1.5cm] (PC5) {Calculate porosity ($\phi$), saturations ($S_i$) and void fraction ($\alpha_v$)};
		\node [block, below of=PC5, node distance=1.4cm] (PC6) {Update the flux $\Phi$ and velocities};
		\node [block1, below of=PC6, node distance=1.2cm] (PC9) {iter $<$ nOuterCorrectors \& residual $>$ tolerance?};
		\node [block2, right of=PC9, node distance=5.0cm] (PC10) {Yes};
		\node[inner sep=0,minimum size=0, node distance=2.0cm,right of=PC10] (PC11) {}; 
		\node [block2, below of=PC9, node distance=1.2cm] (PC12) {No};
		\node [block1, below of=PC12, node distance=1.2cm] (PC13) {End of external iteration};

        \node [noborder, left of=PC1, node distance=4.0cm] (PC1n) {(1)};
        \node [noborder, left of=PC2, node distance=4.0cm] (PC2n) {(2)};
        \node [noborder, left of=PC3, node distance=4.0cm] (PC3n) {(3)};
        \node [noborder, left of=PC4, node distance=4.0cm] (PC4n) {(4)};
        \node [noborder, left of=PC5, node distance=4.0cm] (PC5n) {(5)};
        \node [noborder, left of=PC6, node distance=4.0cm] (PC6n) {(6)};
		
		\path [line] (PC0) -- (PC1);
		\path [line] (PC1) -- (PC2);
		\path [line] (PC2) -- (PC3);
		\path [line] (PC3) -- (PC4);
		\path [line] (PC4) -- (PC5);
		\path [line] (PC5) -- (PC6);
		\path [line] (PC6) -- (PC9);
		\path [line] (PC9) -- (PC10);
		\path [line2] (PC10) -- (PC11);
		\path [line] (PC11) |- (PC1);
		\path [line] (PC9) -- (PC12);
		\path [line] (PC12) -- (PC13);
		
		\end{tikzpicture}
	\end{center}
	\caption{Flowchart for the external iteration in the \texttt{coupledMatrixFoam} solver.}
	\label{fig:flux_cMF}
\end{figure}
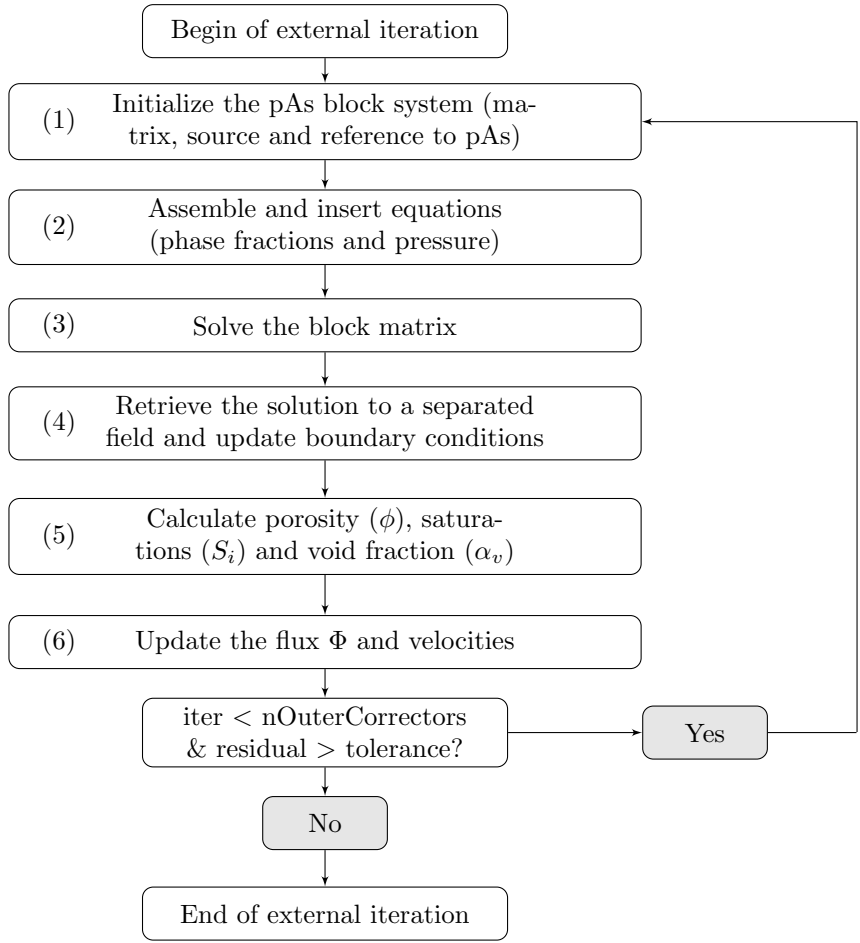

Now, the main operations in the \texttt{coupledMatrixFoam} will be scrutinized in view of clarifying the procedure implemented in the solver. A flowchart with the main steps is shown in \figureautorefname~\ref{fig:flux_cMF}.

The solver is composed of an outer loop that is related to the advance in time and an inner loop that is controlled by a maximum number of iterations or by a tolerance based on the residual related to changes in the pressure solution. All the operations for the linear system solution are done in the inner iteration.

Aiming to help connect the flowchart and the code, the portion related to the inner iteration is shown in code listing \ref{lst:coupledMatrixFoamInnerCode}.

\begin{lstlisting}[
    style=customcpp, 
    caption=Inner iteration of \textit{coupledMatrixFoam.C}, 
    label={lst:coupledMatrixFoamInnerCode}
]
// --- Pressure-alphas corrector loop
while (pAsBlock->loop())
{
    // Initialize the pAs block system (matrix, source and reference to pAs)
    pAsBlock->newMatrix();

    // Assemble and insert equations
    #include "pAsEqn.H"

    // Solve the block matrix
    pAsBlock->solve();

    // Retrieve solution and post-solve
    #include "postSolve.H"
}
\end{lstlisting}

In line 5 of the code listing \ref{lst:coupledMatrixFoamInnerCode} the step (1) of \figureautorefname~\ref{fig:flux_cMF} is done calling the function \texttt{newMatrix()} that is responsible for resetting the block matrix. The step (2) is entirely implemented inside the \texttt{pAsEqn.H}, following the procedure presented in section \ref{sec:blockBuild}. The solution of the block system (step (3)) is done by calling for \texttt{solve()} in line 11 of the code listing \ref{lst:coupledMatrixFoamInnerCode}. Finally, all the operations related to the post-processing of the solution and adequations for the next iteration (steps (4) to (7)) are implemented in \texttt{postSolve.H} aiming to use a comprehensive structure where most of the modifications for new features could be implemented without modifications in the main file of the solver.

\section{Results}
\label{sec:results}

In this section, we demonstrate the use of the \texttt{coupledMatrixFoam} solver through numerical studies. Initially, the convergence behavior and accuracy of the solver are verified for standard porous media problems with analytical solutions. Then, a comparative example designated to illustrate the capacities of the \texttt{coupledMatrixFoam} in relation to segregated strategies is discussed. As a third example, a heterogeneous core application case with significant capillary pressure effects is presented, where we show an improvement in computational efficiency in comparison to the reference solver \texttt{impesFoam}, available in \texttt{porousMultiphaseFoam} toolbox. Finally, the parallel performance of the proposed solver is evaluated, providing an overview of its computational efficiency.

The aforementioned studies can be found in the repository tests, namely:
\begin{enumerate}
    \item[1.] BuckleyLeverett
    \begin{enumerate}
        \item[1.1] homogeneous
        \item[1.2] heterogeneous
        \item[1.3] homogeneousCapillarity
    \end{enumerate}    
    \item[2.] capillarityGravityEquilibrium
    \item[3.] fluidCompressibilty
    \item[4.] coreCase
\end{enumerate}

The numerical experiments presented below consider a Stabilised Bi-Conjugate Gradient (BiCGStab) solver and Crout Incomplete LU preconditioner with zero level of fill-in (ILUC0). A tolerance of $10^{-8}$ is set for the system and $10^{-3}$ for the external loop, that is defined with $\text{nOuterCorrectors}=20$. 

The implicit approximation allows for a time step in principle without limitations, but the Courant condition given by
\begin{equation}
	\text{Co} =  \frac{1}{2}\max_{\alpha_i,\, j} \left( \frac{ \sum_{f=1}^{m_j} |\phi_{f}|}{V_j} \right) \Delta t
	\label{eq:Co_condition}
\end{equation}
will be used as a reference, where $\phi_{f}$ are the total fluxes of phase fraction $\alpha_i$ through each face $f$ of cell $j$, $m_j$ is the total number of faces of cell $j$, $V_j$ is the volume of cell $j$, and $\Delta t$ is the previous time step. In this way, the time step is such that
\begin{equation}
\Delta t_{\max} = \dfrac{\text{Co}_{\max}}{\text{Co}}, 
\label{eq:deltaT}
\end{equation}
where $\text{Co}_{\max}$ is a user-defined value.

\subsection{Verification problems}
\label{subsec:verification}

In the following, we present a verification of the new solver through a series of tests, including the approximation of homogeneous and heterogeneous Buckley-Leverett problems, a capillary-gravity equilibrium case, and a scenario incorporating fluid compressibility effects. This verification follows the steps outlined in \cite{upstream_paper} for the development of a segregated solver.

\subsubsection{Homogeneous Buckley-Leverett}
\label{subsec:BL_homo}

The solver ability to predict two-phase flow in porous media is validated against the well-known semi-analytical Buckley-Leverett solution for viscous-dominated problems, without gravity and capillary pressure effects \cite{wu2015multiphase}. 

A one-dimensional domain of size 0.065\,m fully saturated with oil (denoted by $o$) is considered, where water (denoted by $w$) is injected at a constant rate of $1.35\times10^{-5}$\,m/s on the left side, while the pressure on the right side is fixed at 0.1\,MPa. The Brooks and Corey
relative permeability model with $\eta = 2$ and $k_{r,i(\max)} = 1$ is applied for both phases, whose properties are $\rho_w=1000\,\mbox{kg/m}^{3}$, $\rho_o=800\,\mbox{kg/m}^{3}$, $\mu_w=0.001\, \mbox{Pa}\cdot\mbox{s}$, and $\mu_o=0.002\, \mbox{Pa}\cdot\mbox{s}$. In this example, the absolute permeability is $\mathbf{K}=1\times10^{-13} \, \text{m}^2$, and a homogeneous porosity of $\alpha_v=0.2$ is set. An adjustable time step given by Eq. (\ref{eq:deltaT}) with $\text{Co}_{\max}=1$ has been used.

\begin{figure}[!h]
    \centering
    \includegraphics[width=8cm]{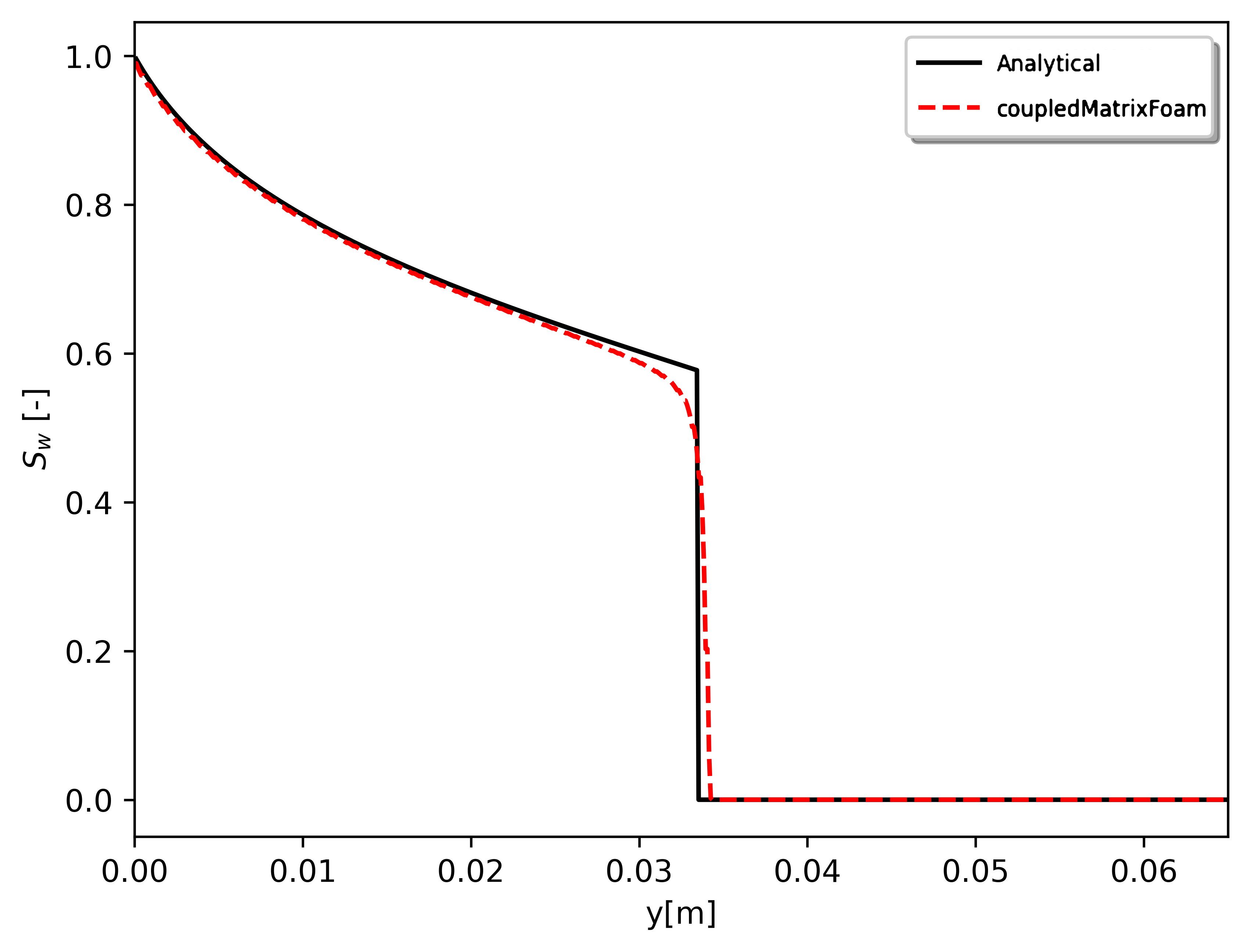}
	\caption{Saturation profile after 1800\,s of the homogeneous Buckley-Leverett case.}
	\label{fig:homogeneous_BL}
\end{figure}

As water flows, it creates a saturation profile with a water shock at its front, as illustrated in Fig. \ref{fig:homogeneous_BL} for a mesh with 500 computational cells at time $t=1800$\,s. Note that the numerical solution obtained agrees with the analytical solution.

This case can be found in the folder \texttt{tests/BuckleyLeverett/homogeneous}.

\subsubsection{Heterogeneous Buckley-Leverett}
\label{subsec:BL_het}
We assess the same system of the previous example, but considering a heterogeneous porosity field such that $\alpha_v=0.2$ if $0\leq x \leq 0.0325$, and $\alpha_v=0.1$ if $0.0325<x \leq 0.065$. 

\begin{figure}[!h]
    \centering
    \includegraphics[width=8cm]{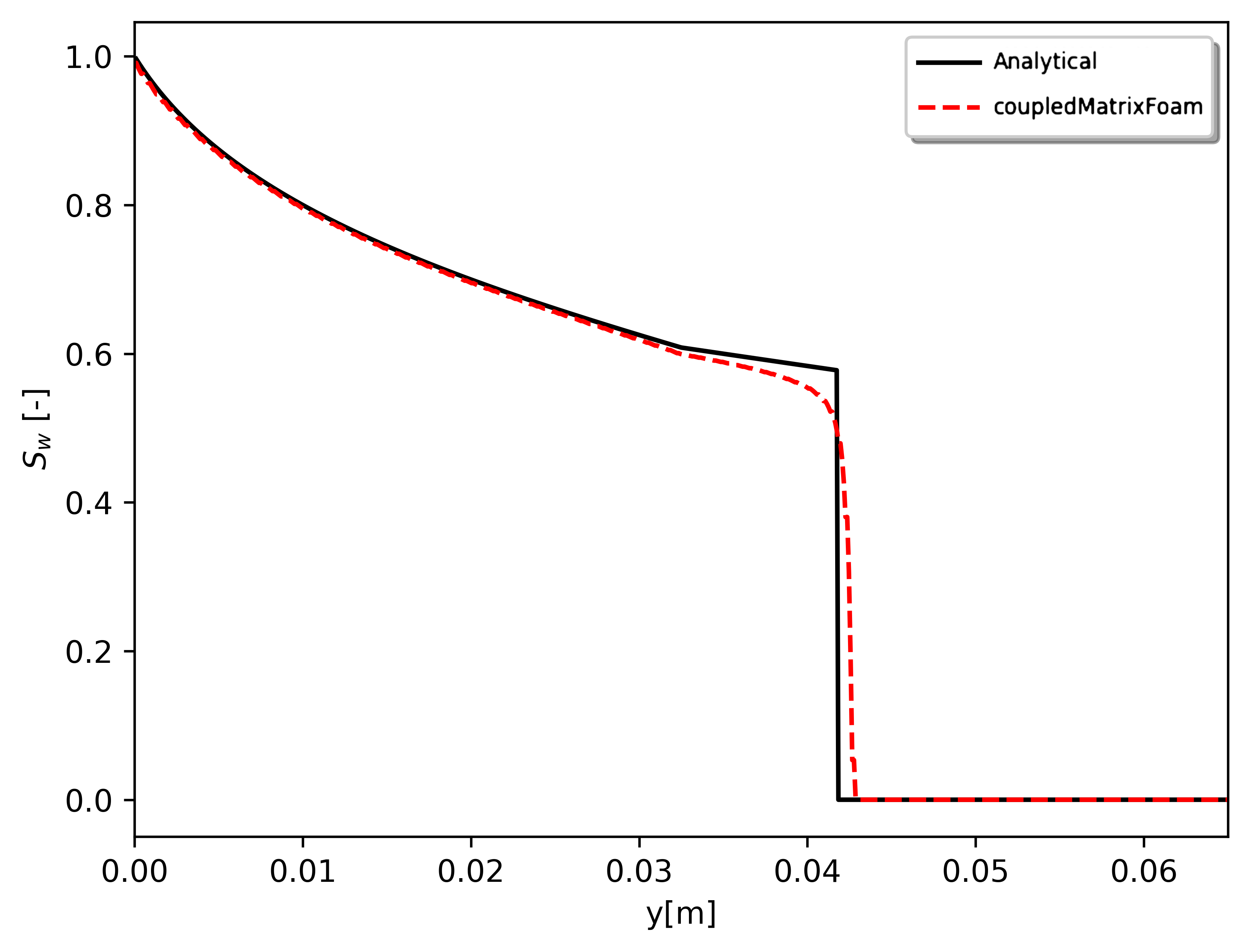}
	\caption{Saturation profile after 1800\,s of the heterogeneous Buckley-Leverett case.}
	\label{fig:heterogeneous_BL}
\end{figure}

The saturation proﬁle is shown in Fig. \ref{fig:heterogeneous_BL}, where we can note a variation in the saturation curve in relation to the case of Fig. \ref{fig:homogeneous_BL} due to the porosity discontinuity. For this example, the numerical solution obtained also agrees with the analytical solution.

This case can be found in the folder \texttt{tests/BuckleyLeverett/heterogeneous}.

\subsubsection{Capillary-gravity equilibrium}
\label{subsec:pc_gravity}
This example aims to evaluate the solution in the presence of gravity and capillary pressure. An air-water flow in a one meter vertical domain is carried out until the gravity-capillarity equilibrium. 

The initial condition considers the lower half containing water ($S_w = 0.5$) and air ($S_a = 0.5$), while the upper half contains only air ($S_w = 0$ and $S_a = 1$). At the bottom boundary, a Neumann zero condition is applied, while the top boundary receives a Dirichlet fixed pressure of 1 MPa. Under these conditions, the water saturation at the theoretical steady state is such that:
\begin{equation}
\frac{\partial S_w}{\partial y}=\frac{(\rho_a-\rho_w)g_y}{\frac{\partial p_c}{\partial S_w}},
\label{eq:derivada_sat_equilibrio}
\end{equation}
where $g_y=-9.8\,\mbox{m/s}^{2}$ is the gravity acceleration \cite{horgue2015open, carrillo2020multiphase}. The capillary pressure model used is the Brooks and Corey \cite{brooks1965hydraulic}, which is given by 
\begin{equation}
p_c=p_{c,0}\,\left(\frac{\alpha_i}{\alpha_v}\right)^{-\beta},
\label{eq:pc_BC}
\end{equation}
where we set $p_{c,0}=1000$\,Pa and $\beta=0.5$. 

The Brooks and Corey relative permeability model with $n=3$ and $k_{r,i(\max)}=1$ is considered for both air and water. Other settings include homogeneous porosity $\alpha_v=0.5$, absolute permeability $\mathbf{K}=1\times 10^{-11}\,\mbox{m}^2$, and phase properties $\rho_w=1000\,\mbox{kg/m}^{3}$, $\rho_a=1\,\mbox{kg/m}^{3}$, $\mu_w=0.001\, \mbox{Pa}\cdot\mbox{s}$, and $\mu_a=1.76\times 10^{-5}\, \mbox{Pa}\cdot\mbox{s}$. 

\begin{figure}[!h]
  \centering
    \begin{subfigure}[a]{0.47\textwidth}
         \centering
         \includegraphics[width=\textwidth]{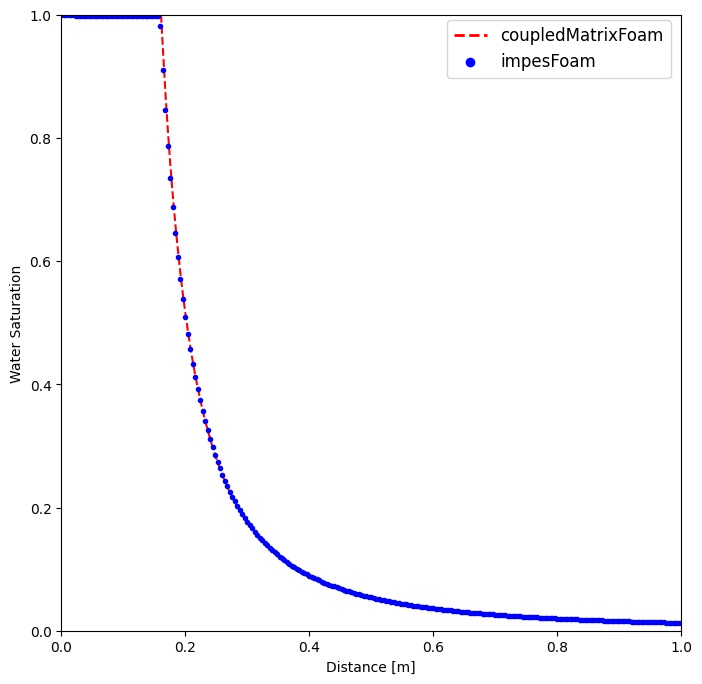}
         \caption{Saturation profile.}
          \label{fig:homogeneous_pc_gravity_a}
     \end{subfigure}
    \begin{subfigure}[a]{0.47\textwidth}
         \centering
         \includegraphics[width=\textwidth]{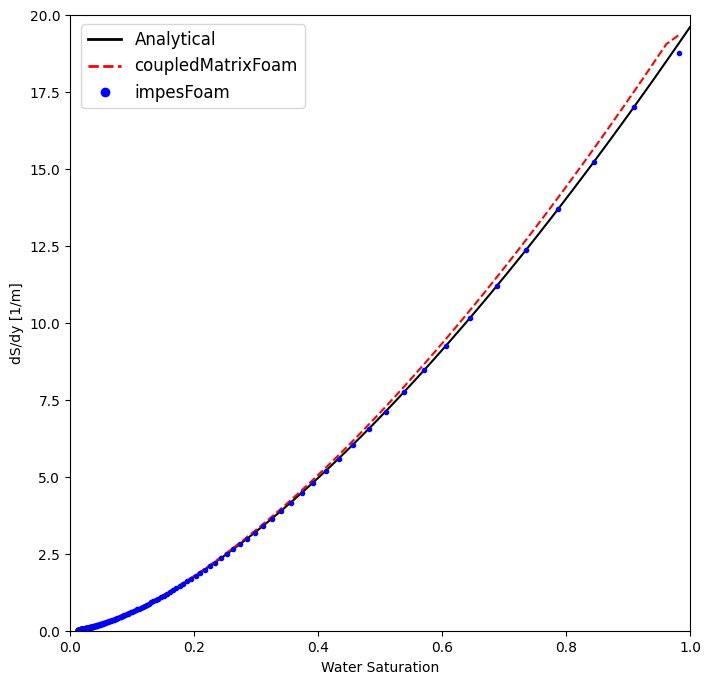}
         \caption{Saturation gradient.}
          \label{fig:homogeneous_pc_gravity_b}
     \end{subfigure}  
  \caption{Capillary-gravity equilibrium test.}
\end{figure}

A final time of $115.74$\,days has been chosen in order to ensure that the capillary-gravity equilibrium is attained. In this example, the time step was chosen starting at 1\,s increasing by 20\% until reaching 100\,s. Figure \ref{fig:homogeneous_pc_gravity_a} shows the water saturation profile obtained from \texttt{coupledMatrixFoam} for a 1000 cells grid compared with the \texttt{impesFoam} solution for the same problem. We remark that this example is a validation case for the \texttt{impesFoam} solver presented in \cite{horgue2015open}, where Toddy time step control is used. Note that the \texttt{coupledMatrixFoam} and \texttt{impesFoam} solutions are comparable. The water saturation gradients are presented in Fig. \ref{fig:homogeneous_pc_gravity_b}, both in agreement with the analytical curve.

This case can be found in \texttt{tests/capillarityGravityEquilibrium}.

\subsubsection{Fluid compressibility effects}
\label{subsec:compressibility}
In this experiment, we investigate a compressible air flow through porous media by assessing mass conservation under varying time-step size.

A one-dimensional domain of 0.065\,m in length is fully saturated with air and discretized into 1000 cells. Due to fluid compressibility, a transient production occurs at the right boundary when the system, initially at 1\,MPa, is subjected to a fixed pressure of 0.1\,MPa at that boundary. The setup for this problem considers an air viscosity of $\mu=1.716\times 10^{-5}\, \mbox{Pa}\cdot\mbox{s}$, absolute permeability of $\mathbf{K}=1\times10^{-15} \, \text{m}^2$, porosity of $\alpha_v=0.2$, and neglects gravity effects.

\begin{figure}[!h]
    \centering
    \includegraphics[width=10cm]{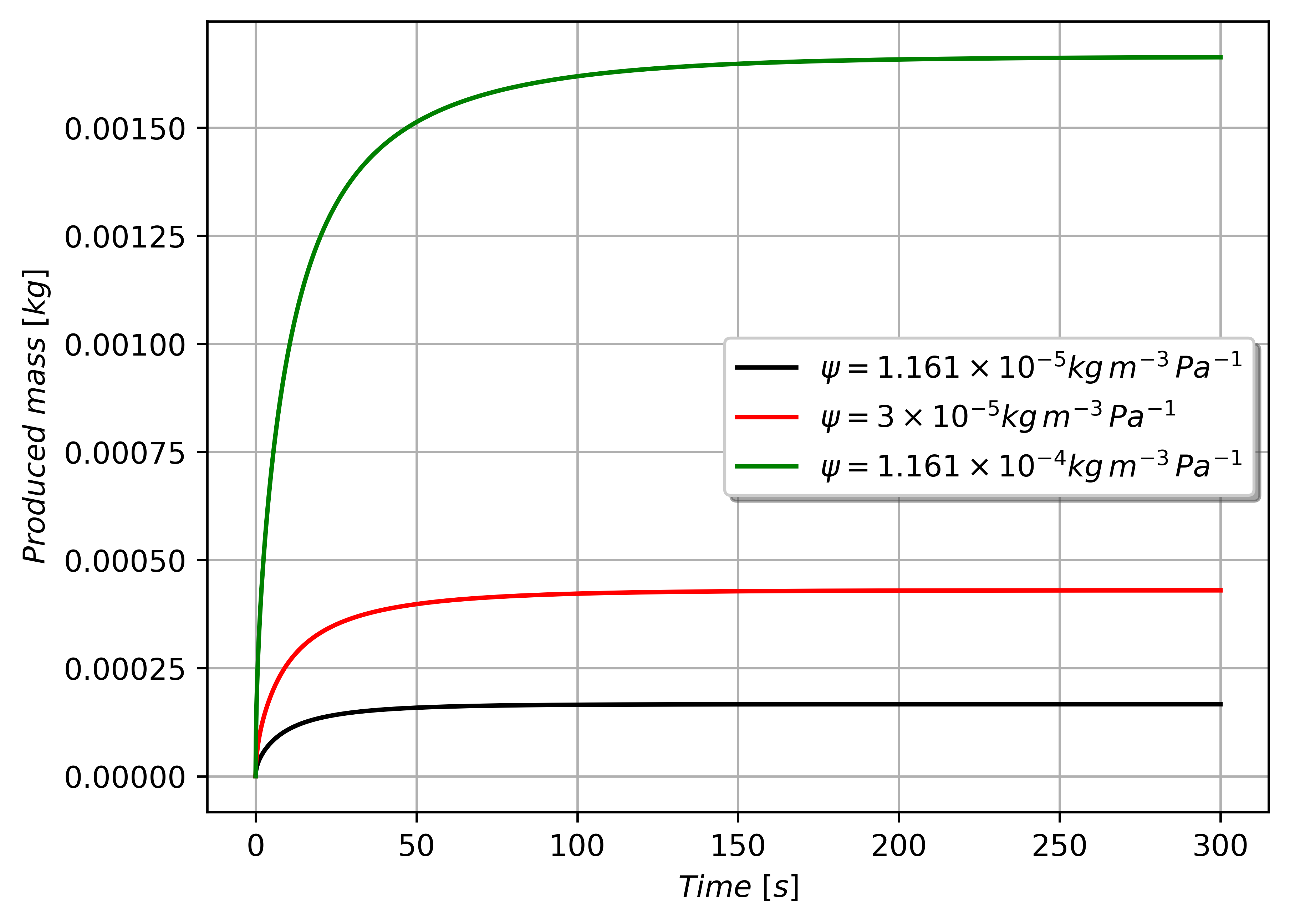}
	\caption{Air produced mass for three different compressibility factors.}
	\label{fig:compressible}
\end{figure}

The propagation of air through the porous medium leads to a pressure field that evolves over time, resulting in gradual pressure changes and corresponding variations on mass produced depending on the compressibility factor. Figure \ref{fig:compressible} compares the mass produced for three different compressibility factors, using a fixed time-step size of $\Delta t=0.1\,$s and a final simulation time of 300\,s. As expected, mass production increases significantly with higher compressibility.

Using the compressibility model given by Eq. (\ref{eq:compressibility}), it is possible to estimate the air produced mass by
\begin{equation}
    m= \alpha_v\,V\,(\rho^i-\rho^f)=\alpha_v\,V\,\psi\,(p^i-p^f),
\end{equation}
where $\rho^i$ and $\rho^f$ are the initial and final densities, respectively, and $p^i$ and $p^f$ denote the initial and final pressures. The reference density considered in the cases reported in Fig. \ref{fig:compressible} is $\rho_0=1.292$\,kg/m$^3$.

At final time, it is assumed that $p^f$ is equal to the pressure imposed at the outlet boundary. Therefore, for the compressibility factor of $\psi=1.161\times 10^{-5}$\,kg\,m$^{-3}$\,Pa$^{-1}$, the analytical produced mass is $m=0.0001664$\, kg, which corresponds to a variation on the density from $\rho^i=12.902$\,kg/m$^3$ to $\rho^f=2.453$\,kg/m$^3$. To evaluate the corresponding mass conservation of air during this simulation, we tested different values of time-step size. The results are shown in Table \ref{tab:head_loss}, where the agreement between the numerical and analytical mass values confirms that the solver accurately enforces mass conservation in all tested time-step sizes. Note that for the last case, the simulation used only three time steps and still preserved mass conservation.
\begin{table}
\centering
  \begin{tabular}{ccccc}
    \hline 
    $\Delta t$ [s] & $m$ [kg] & Error [kg]  \\
    \hline 
	0.1  & 0.000166399 &  $1.325\times 10^{-9}$ \\ 
	1    & 0.000166399 &  $1.325\times 10^{-9}$ \\ 
	10   & 0.000166395 & $5.325\times 10^{-9}$ \\  
	100  &  0.000165404 & $9.9632\times 10^{-7}$ \\ \hline 
  \end{tabular}
\caption{The mass of air produced as a function of the time-step size.}
\label{tab:head_loss}
\end{table}

This case can be found in \texttt{tests/fluidCompressibility}.

\subsection{Stability analysis}
\label{subsec:comparison}

In order to investigate the stability of the proposed method, a temporal convergence for the homogeneous Buckley-Leverett problem adding capillarity effects is presented. We consider the reference solution given by a grid with 500 computational cells and fixed time-step $\Delta t=10^{-4}$\,s and study the behavior of approximations for different time refinements. The Brooks and Corey capillarity model is used with $p_{c,0}=10$\,Pa and $\beta=0.5$.

\begin{figure}[!h]
  \centering
    \begin{subfigure}[a]{0.49\textwidth}
         \centering
         \includegraphics[width=\textwidth]{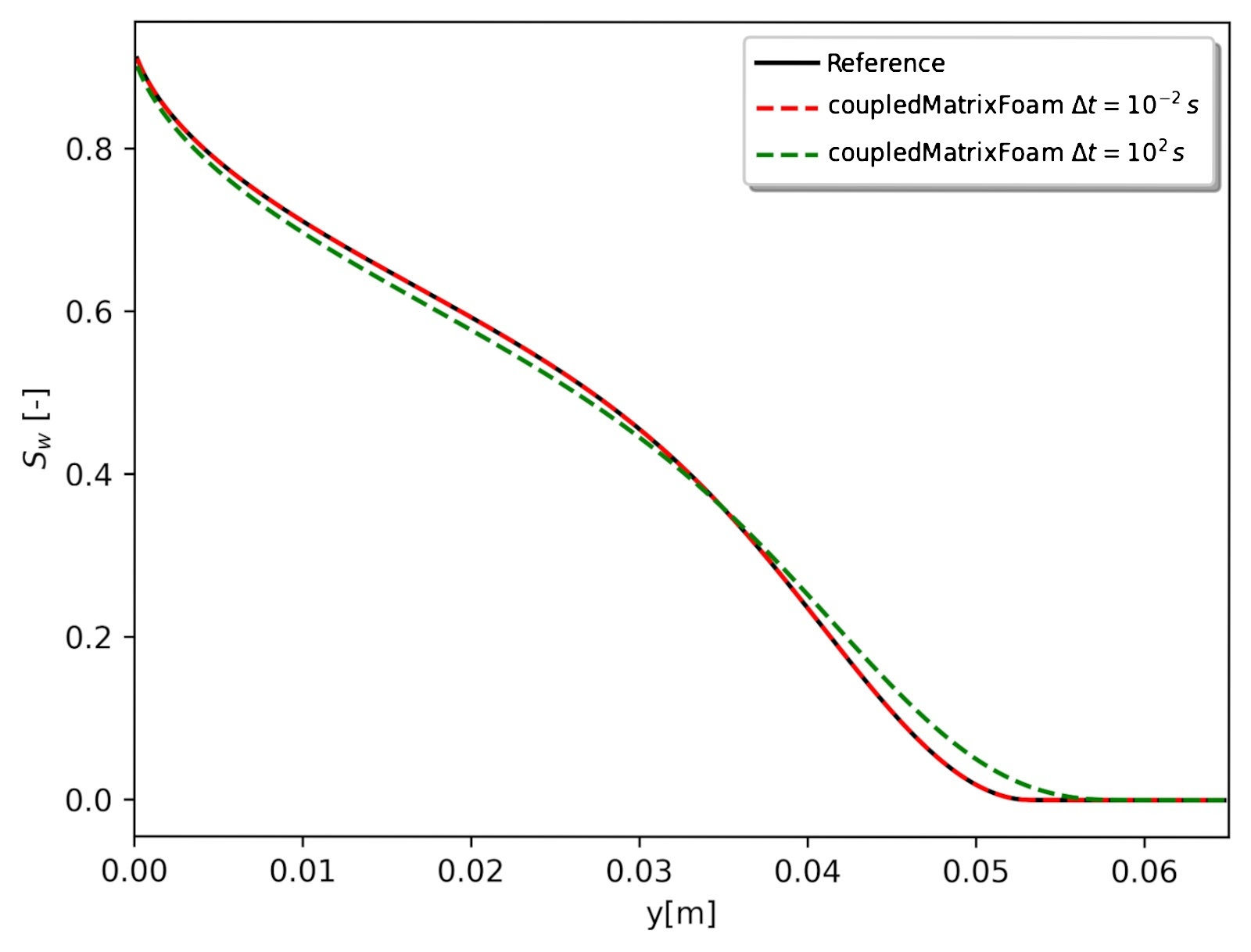}
         \caption{Saturation profiles.}
          \label{fig:stability_analysis}
     \end{subfigure}
    \begin{subfigure}[a]{0.49\textwidth}
         \centering
         \includegraphics[width=\textwidth]{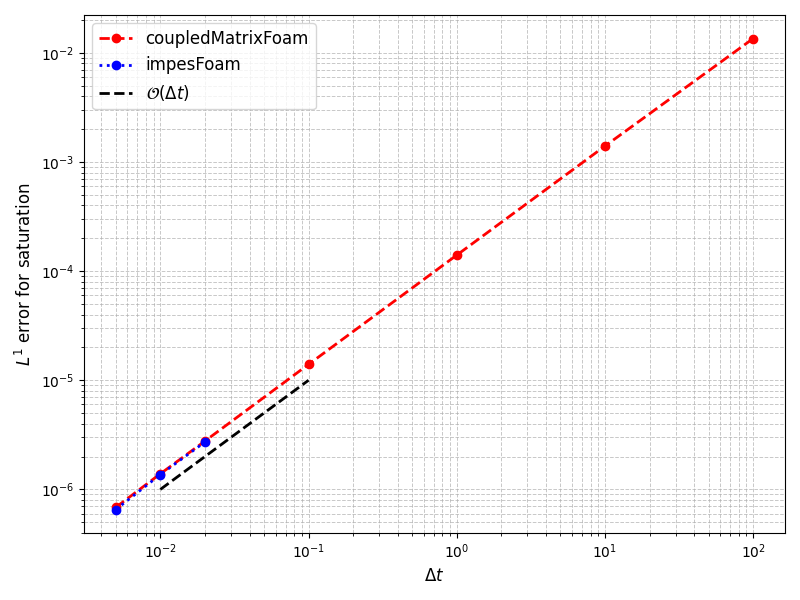}
         \caption{$L^1$ errors.}
          \label{fig:time_convergence}
     \end{subfigure}  
  \caption{Saturation as a function of the time step size.}
\end{figure}

When using the fully implicit strategy, the choice of time step size depends only on the accuracy request. Figure \ref{fig:stability_analysis} presents the \texttt{coupledMatrixFoam} approximations for $\Delta t=10^{-2}$\,s and $\Delta t=10^{2}$\,s, where it is clearly noted that the solutions are more accurate for the smaller time interval. In Fig. \ref{fig:time_convergence}, we can confirm the ability of the \texttt{coupledMatrixFoam} solver to increase the time step when compared to the explicit method applied by the \texttt{impesFoam}, which has a time step restricted by the CFL (Courant-Friedrichs-Lewy) condition. Note that both solvers returned the expected linear slope, with the \texttt{coupledMatrixFoam} being confirmed as an excellent alternative to reduce the computational cost of simulations.

This case can be found in \texttt{tests/BuckleyLeverett/homogeneousCapillarity}.

\subsection{Application case}
\label{subsec:application}
\begin{figure}[!h]
    \centering
    \includegraphics[width=13cm]{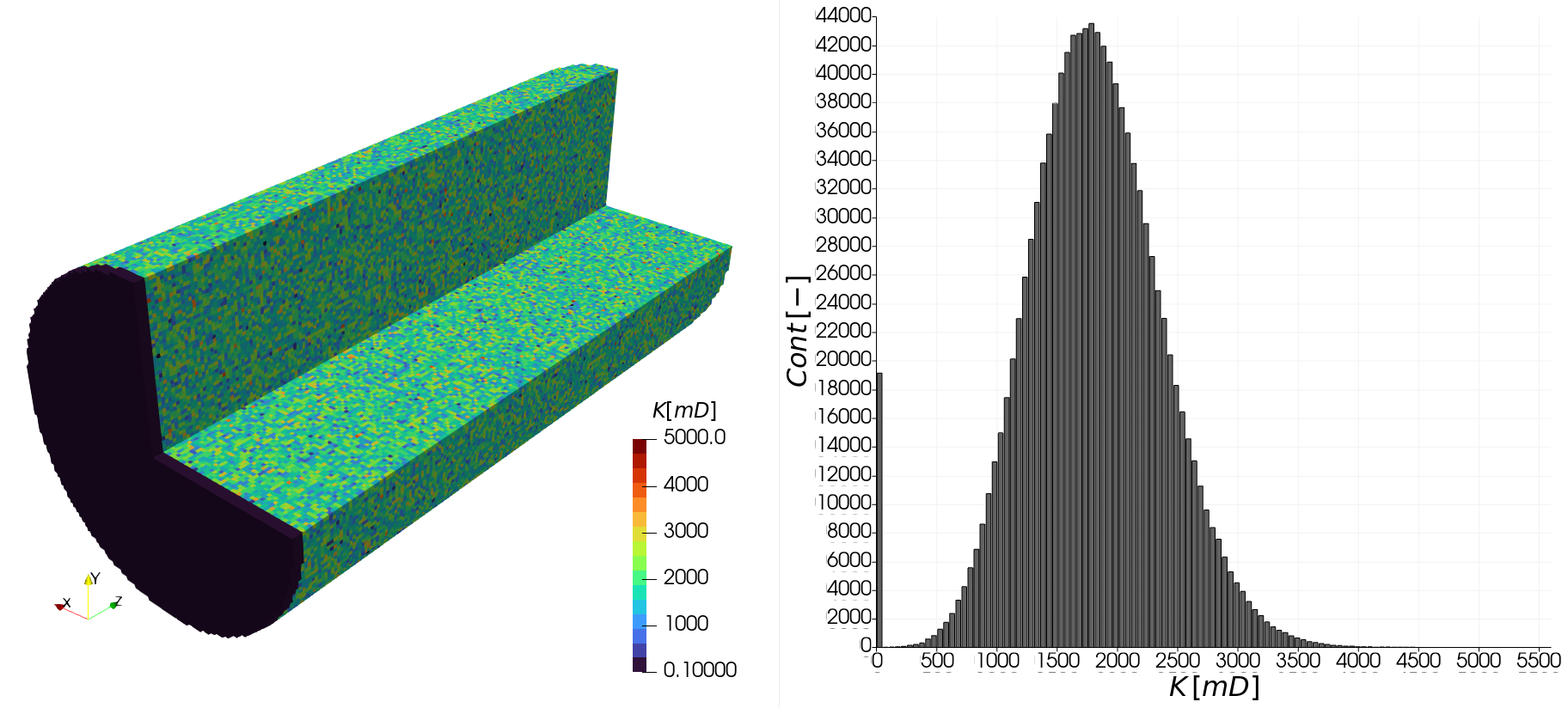}
	\caption{Randomly generated absolute permeability field.}
	\label{fig:plug_perm}
\end{figure}

The last experiment presents a simulation of an oil-water flow in a 3D heterogeneous core case, where the possibility of increasing the time step is very significant in terms of computational cost. The heterogeneous permeability field randomly generated is illustrated in Figure \ref{fig:plug_perm}, while the porosity field is constant $\alpha_v = 0.2$.

The computational domain with 1.17\,M cells distributed in $0.09\,\text{m}\times 0.09\,\text{m}\times 0.183\,\text{m}$ is initially saturated with oil and water is injected from left at a constant rate of $1.31 \times 10^{-5}$m/s, with the pressure at right fixed in 101.35\,MPa.  The Brooks and Corey relative permeability model with $\eta = 2$ and $k_{r,i(\max)} = 1$ is applied for both phases, whose properties are $\rho_w=1000\,\mbox{kg/m}^{3}$, $\rho_o=800\,\mbox{kg/m}^{3}$, $\mu_w=0.001\, \mbox{Pa}\cdot\mbox{s}$, and $\mu_o=0.002\, \mbox{Pa}\cdot\mbox{s}$. We also consider the Brooks and Corey model for capillary pressure with $\beta=0.5$ and $p_{c,0}=100$\,Pa, while gravity is $g_y=-9.8\,\mbox{m/s}^{2}$.

\begin{figure}[!h]
    \centering
    \includegraphics[width=13cm]{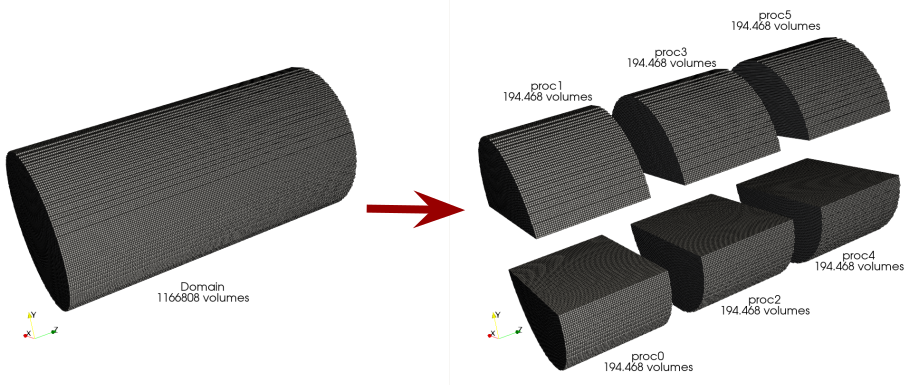}
	\caption{Domain partition scheme for parallel solution.}
	\label{fig:plug_partition}
\end{figure}

This application features a large mesh, so both \texttt{coupledMatrixFoam} and \texttt{impesFoam} were run in parallel on a 12th Gen Intel i7-12700H CPU using six processors, as illustrated in Fig. \ref{fig:plug_partition}. The domain decomposition was performed with the simple method, creating three partitions along the Z-axis and two along the Y-axis. No special MPI flags were applied, and the CPU did not run in exclusive mode.

\begin{figure}[!h]
    \centering
    \includegraphics[width=8cm]{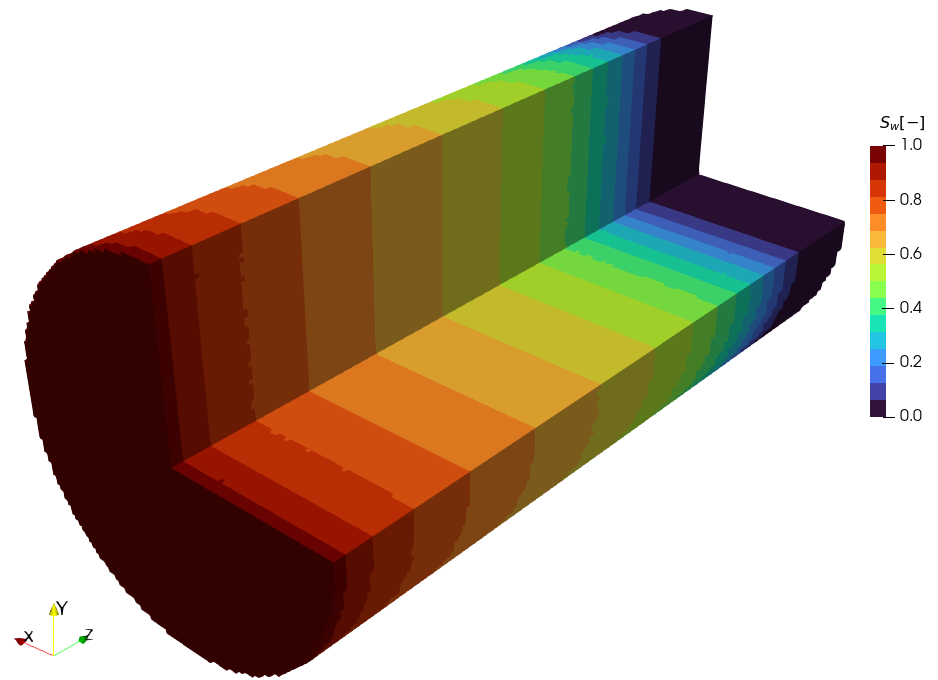}
	\caption{Saturation profile at time $t=2$\,h.}
	\label{fig:plug_sat}
\end{figure}

\begin{figure}[!h]
  \centering
    \begin{subfigure}[a]{0.43\textwidth}
         \centering
         \includegraphics[width=\textwidth]{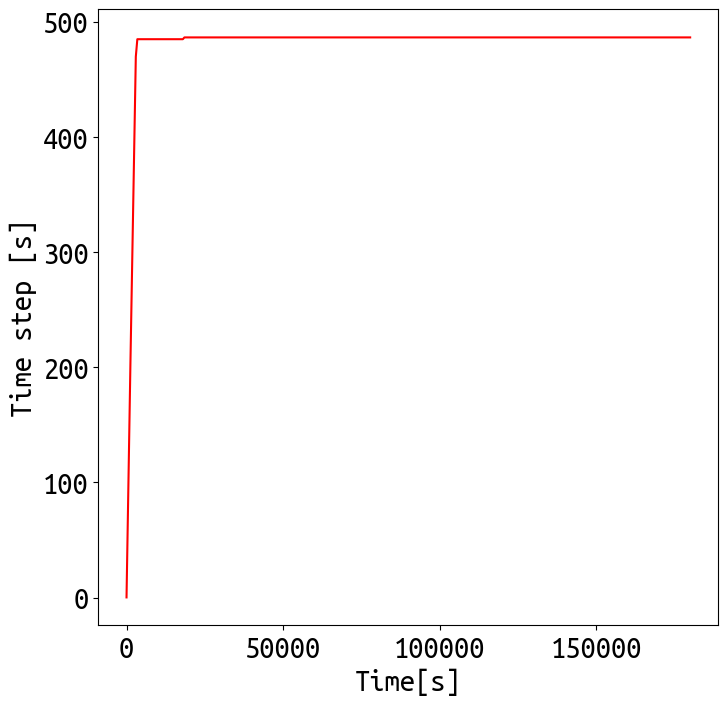}
         \caption{\texttt{coupledMatrixFoam}.}
     \end{subfigure}
    \begin{subfigure}[a]{0.43\textwidth}
         \centering
         \includegraphics[width=\textwidth]{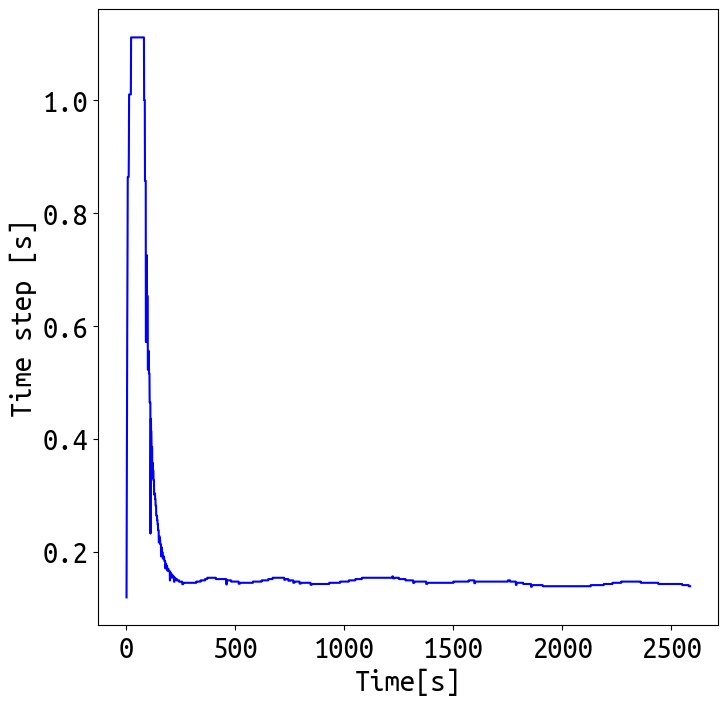}
         \caption{\texttt{impesFoam}.}          
     \end{subfigure}  
  \caption{Time step histories.}
       \label{fig:time_steps}
\end{figure}

In this example, the effect of capillary pressure is very significant and produces a restricted time step size when using explicit methods. An adjustable time step given by Eq. (\ref{eq:deltaT}) with $\text{Co}_{\max}=33$ has been used for the \texttt{coupledMatrixFoam} solver. Figure \ref{fig:plug_sat} illustrates the saturation solution at time $t=2$\,h. The final time has been set at 50h, requiring a computational execution time of 1.49h. When running \texttt{impesFoam} under equivalent conditions with $\text{Co}_{\max}=0.75$, it was possible to reach 0.72\,h of simulation after 13.02\,h of computational execution time. Therefore, it is expected from \texttt{impesFoam} a running time of 904.17\,h to run the equivalent final time of \texttt{coupledMatrixFoam}. Respective time step histories are presented in Fig \ref{fig:time_steps} where it is possible to observe that the time step stabilizes around 0.2 seconds for \texttt{impesFoam} while for the \texttt{coupledMatrixFoam} this stabilizes around 500 seconds. Therefore, an increase on the order of 2,500 for the time step has been achieved. This result reinforces the ability of the implicit method to abruptly reduce the computational cost.

This case can be found in the folder \texttt{tests/coreCase}.

\subsection{Parallel performance evaluation}
\label{subsec:speedup}

The \texttt{coupledMatrixFoam} solver was developed with industrial applications in mind. Since such cases typically run on high-performance computing (HPC) resources, the final set of results evaluates the performance of the application case (Section \ref{subsec:application}) in parallel, using different processor counts.

As mentioned in Section \ref{subsec:application}, the baseline simulation used six processors. For the speed-up tests, an AMD EPYC 7742 CPU (64 cores) was employed in exclusive mode, without hyperthreading, scaling from a single core (serial execution) to all 64 cores. In these tests, the domain decomposition was carried out using the \textit{scotch} method. 
To measure performance, the simulation started from a previous time of 10,000 seconds, then ran for an additional 1,000 seconds of physical time with a fixed time step of 5, seconds, resulting in 200 iterations. Furthermore, the solver was configured to perform 20 iterations of the coupled linear system every running two internal iterations.

Because the memory requirements of block-system solvers differ substantially from those of segregated solvers like \texttt{impesFoam}, we chose \texttt{pUCoupledFoam} \cite{uroic2019implicitly} as the basis for comparison. This solver, which is natively available in foam-extend for pressure–velocity coupling, was tested using a 2D cavity tutorial case initiated from a previously converged solution. Specifically, the simulation proceeded for 20 seconds—20 iterations at a fixed time step of 1 second - on a mesh of 
1,000 × 1,000 cells, yielding one million total cells. The domain decomposition was performed with the \textit{scotch} method, and the block-system solver executed one internal iteration with 500 iterations per time step.

\begin{figure}[!h]
  \centering
    \begin{subfigure}[a]{0.49\textwidth}
         \centering
         \includegraphics[width=\textwidth]{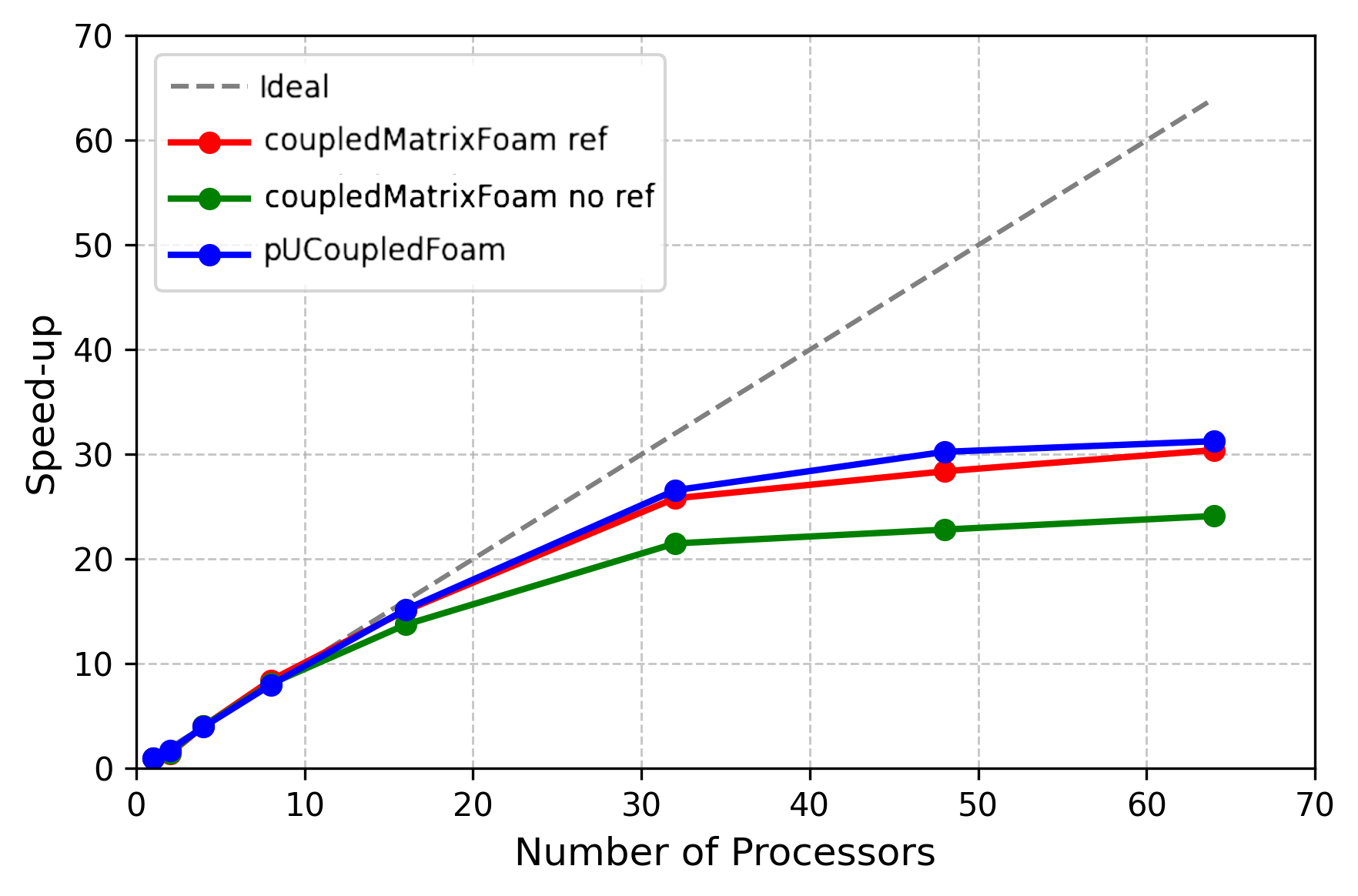}
         \caption{Speed-up.}
          \label{fig:speedup}
     \end{subfigure}
    \begin{subfigure}[a]{0.49\textwidth}
         \centering
         \includegraphics[width=\textwidth]{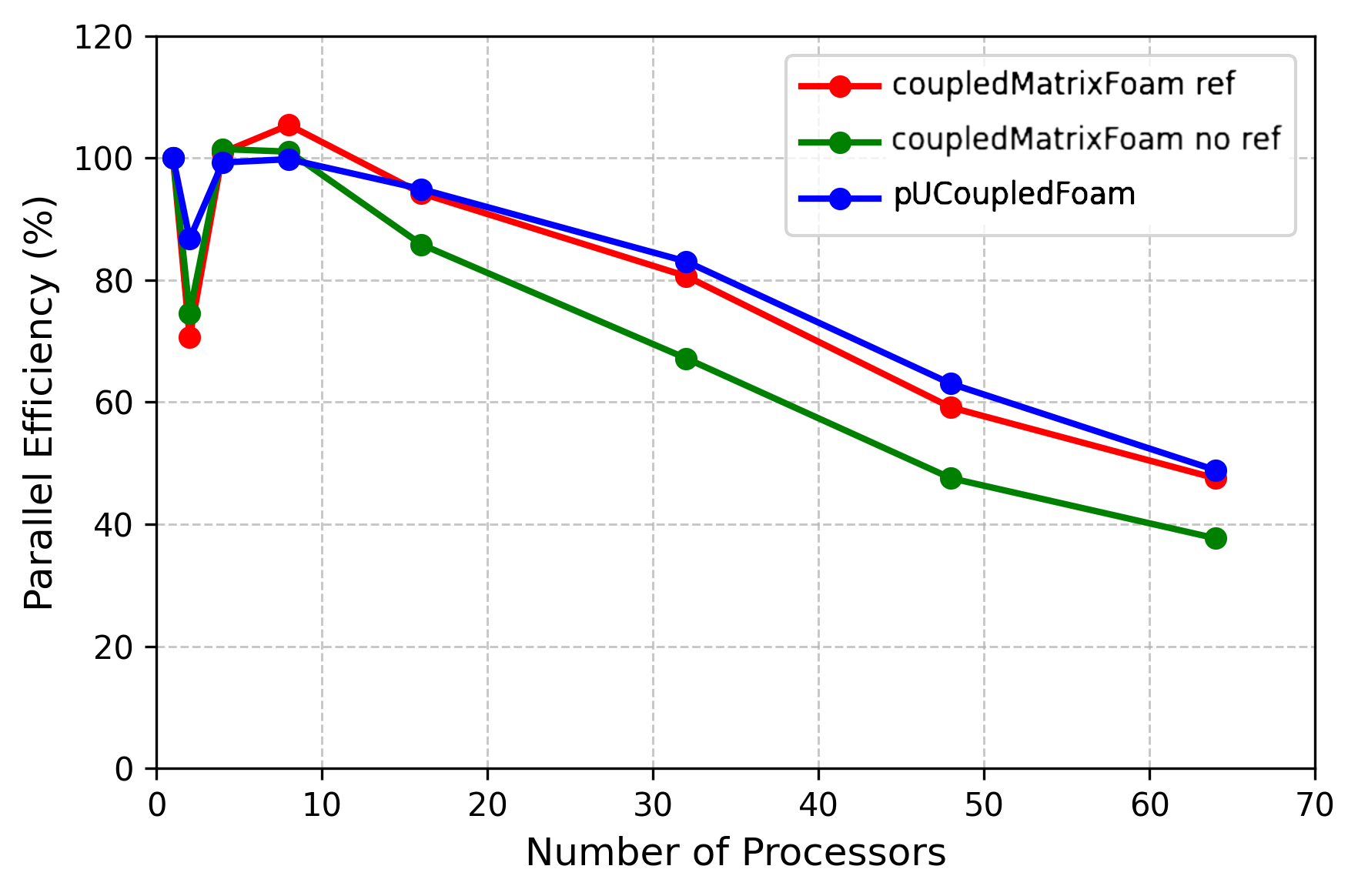}
         \caption{Efficiency [\%].}
          \label{fig:efficiency}
     \end{subfigure}  
  \caption{Comparison of performance for \texttt{coupledMatrixFoam} with and without reference phase with \texttt{pUCoupledFoam}.}
  \label{fig:speedUpResults}
\end{figure}

Several key differences distinguish the 2D tutorial from the main application: the latter is 3D, has about 17\% more cells, and involves either three coupled variables (no reference phase) or two coupled variables (with a reference phase), whereas \texttt{pUCoupledFoam} couples four variables (pressure plus three velocity components). Bearing these differences in mind, the speed-up and efficiency results are presented in Figure~\ref{fig:speedUpResults}.

Figure \ref{fig:speedup} presents the speed-up curves for the three cases - \texttt{coupledMatrixFoam} with (\texttt{coupledMatrixFoam} ref) and without (\texttt{coupledMatrixFoam} no ref) reference phase, and \texttt{pUCoupledFoam} - compared against the ideal linear speed-up. The ideal line represents the theoretical scenario where doubling the number of processors cuts the runtime in half. In Fig. \ref{fig:efficiency} it is shown the parallel efficiency of the three solvers. Efficiency is computed as the ratio of actual speed-up to the ideal speed-up, expressed as a percentage. A value of 100\% corresponds to perfect scaling.

Regarding the scalability of the solver, the speed-up is close to the ideal line for the three solvers until 16 processors present efficiencies higher than 80\% except for the \texttt{coupledMatrixFoam} using two processors. This behavior for two processors could be related to the difference of requirements when comparing serial and parallel simulations. Up to 16 processors the speed-up is no closer to the ideal line but it is interesting to observe that the speed-up and efficiency of \texttt{coupledMatrixFoam} using reference phase is very close to the results of \texttt{pUCoupledFoam}. The behavior using more than 16 processors is expected and discussed in different works \cite{galeazzo2024performance, galeazzo2024understanding} but without considering coupled solvers.

\begin{figure}[!h]
    \centering
    \includegraphics[width=8cm]{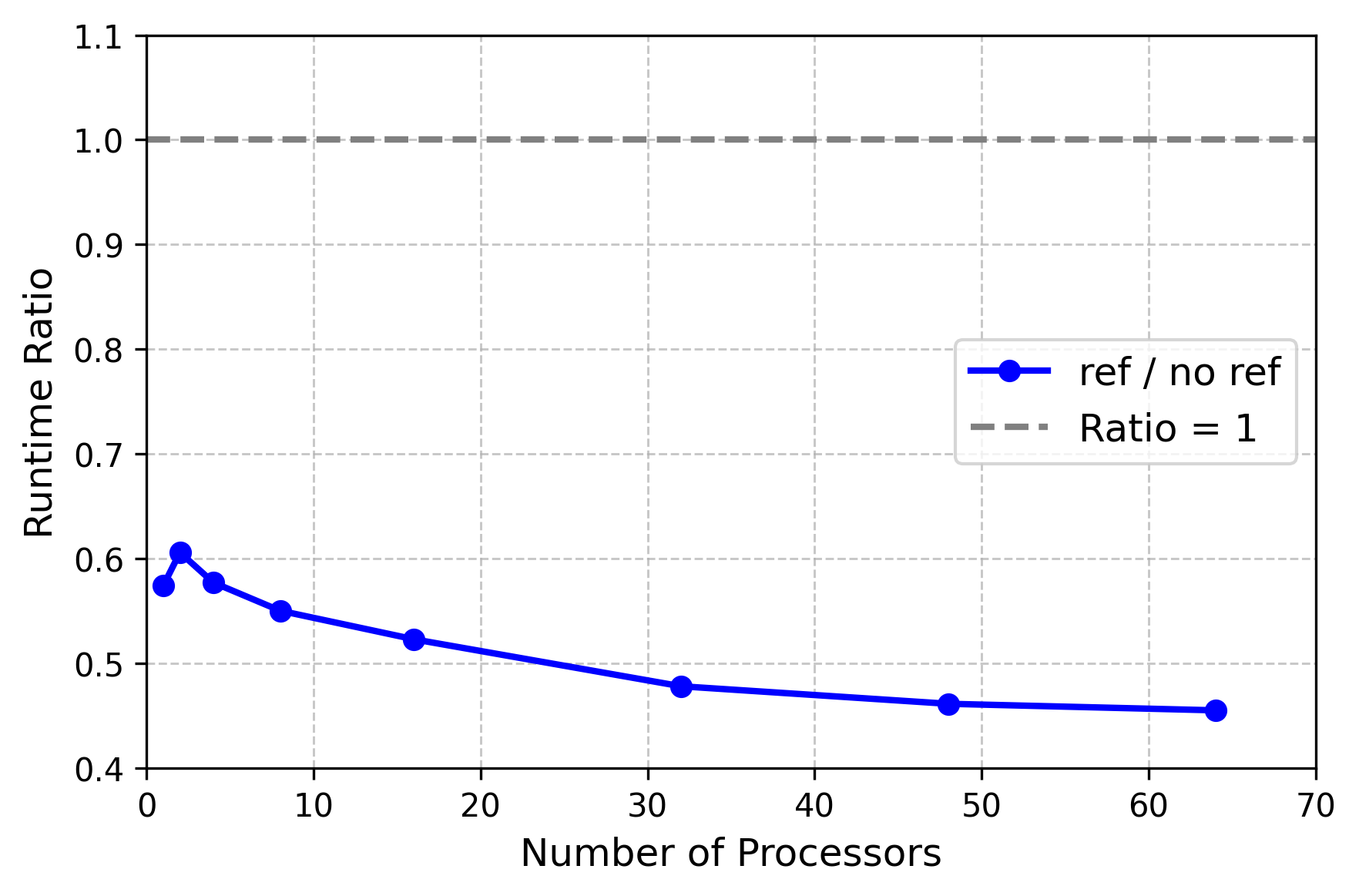}
	\caption{Runtime ratio between \texttt{coupledMatrixFoam} using reference phase over not using.}
	\label{fig:timeRatio}
\end{figure}

Finally, Figure~\ref{fig:timeRatio} specifically compares \texttt{coupledMatrixFoam} runtimes between the scenarios with and without a reference phase. The ratio is defined as
\begin{equation}
    \text{ratio} =
    \dfrac{\text{runtime (reference phase)}}
    {\text{runtime (no reference phase)}}
\end{equation}
and a value below 1 indicates faster execution when using a reference phase.

Up to 30 processors, the runtime is less than half when the reference phase is included, with the largest ratio (approximately 0.6) observed at two processors. Effectively, removing one phase from the block system nearly doubles the solver speed, highlighting the complexity involved in analyzing block-system solver performance. Overall, these results provide a baseline for understanding how \texttt{coupledMatrixFoam} behaves under varying configurations and computational resources.

\section{Conclusion}
\label{sec:conclusion}

In this study, \texttt{coupledMatrixFoam} is introduced. It is a new fully coupled, implicit solver 
for simulating multiphase flows in heterogeneous porous media. Developed as part of the 
\texttt{porousMedia} open-source repository maintained by WIKKI Brasil and based on 
\texttt{foam-extend~5.0}, this solver integrates the Eulerian multi-fluid approach with 
Darcy’s law to capture interactions among multiple fluids in complex geological formations 
where porosity and permeability vary significantly.

A key innovation of \texttt{coupledMatrixFoam} lies in its implicit coupling of pressure 
and phase fractions, using a single block matrix system. This approach increases the numerical stability and accelerates convergence when compared to traditional segregated solvers. Comparisons with \texttt{impesFoam} revealed time-step increases on the order of 2,500 for a three-dimensional heterogeneous permeability case, especially under strong 
capillary pressure conditions.

Validation efforts covered well-established benchmarks such as the Buckley--Leverett problem (both homogeneous and heterogeneous cases), a capillary--gravity equilibrium test, and a fluid compressibility case, all of which confirmed the solver’s accuracy. In addition, performance tests on a complex 3D 
scenario underscored the method’s scalability and the effectiveness of the reference phase 
strategy, wherein one fluid phase is treated outside the block system to reduce computational 
demands.

Increasingly the robustness and stability of numerical simulations for multiphase interactions in porous media, \texttt{coupledMatrixFoam} expands the range of possibilities for modeling applications in oil recovery, CO$_2$ storage, and hydrogeology. Future development plans 
include extending the solver to incorporate multi-component flow, reactive transport, and thermal effects, further enhancing its utility in real-world settings.

\section*{Acknowledgements}

\noindent The authors would like to thank Petrobras (grant 2017/00610-4) for the financial and material support.

\section*{Declaration of generative AI and AI-assisted technologies in the writing process }

During the preparation of this work the authors used ChatGPT, a language model developed by OpenAI, in order to improve language and readability. After using this tool/service, the authors reviewed and edited the content as needed and take full responsibility for the content of the publication.



\bibliographystyle{elsarticle-num}
\bibliography{Bibliography}







\end{document}